\documentclass[preprint]{aastex}

\newcommand{\be}{\begin{enumerate}}
\newcommand{\ee}{\end{enumerate}}
\newcommand{\dg}{$^{\circ}$}
\newcommand{\um}{$\mu$m}
\newcommand{\rsun}{R_{\odot}}
\newcommand{\msun}{M_{\odot}}

\usepackage{verbatim}

\usepackage{natbib}

\shorttitle{Variable Stars Near the Galactic Center}
\shortauthors{Peeples, Stanek, \& DePoy}

\title{A Study of Stellar Photometric Variability Within the Central 4~pc
of the Galactic Center with Infrared Image Subtraction\altaffilmark{1}}

\author{Molly~S.~Peeples\altaffilmark{2}, K.~Z.~Stanek\altaffilmark{2}, and
  D.\ L.\ DePoy\altaffilmark{2}}

\altaffiltext{1}{Based on observations from the CTIO/Yale 1-m telescope
with the ANDICAM imager} 
\altaffiltext{2}{Department of Astronomy, Ohio State University, 140 W.\
18th Ave., Columbus,~OH~43210} \email{molly, kstanek,
depoy@astronomy.ohio-state.edu}

\begin{document}

\begin{abstract}
We present a catalog of 110 variable stars within $\sim 1'$ of Sgr~A*
based on image subtraction of near-infrared ($H$ and $K$) photometry.
Our images were obtained over 133 nights from 2000 to 2002 in $H$-band
and over 134 nights from 2001 to 2002 in $K$-band; the typical FWHM is
$1.4''$.  We match the catalog to other near-infrared, X-ray, and radio
(i.e., maser) data, and we discuss some of the more interesting objects.
The catalog includes 14 sources with measurable periods, several known
long-period variables and three new LPV candidates.  We associate
IRS~10* with OH, SiO, and H$_2$O masers and a bright X-ray point source;
this analysis suggests IRS~10* is an AGB star with an accreting
companion.  Among the $\approx 90$ newly discovered sources are a
probable cataclysmic variable, a potential edge-on contact 84~day period
eclipsing binary, and a possible 41~day period pulsating variable.
\end{abstract}


\section{Introduction}\label{sec:intro}

The Galactic center offers a unique opportunity for studying star
formation and stellar populations near a supermassive black hole (SMBH),
namely, Sgr~A*.  Active galactic nuclei and supernovae feedback have
been evoked to explain many of the observed inhomogeneities in different
galaxies \citep[e.g.,][]{silk97,efstathiou00,scannapieco05}, but it is
only in our own Galaxy that we are able to resolve individual stars
within the circumnuclear ring surrounding the central SMBH
\citep{jackson93}.  Within the Milky Way, the cluster of stars at the
Galactic center is also unique in that it is comprised of both old and
young stars \citep[see e.g.,][]{lacy82, paumard06}.  It is believed that
the young stellar population is the result of a burst of star formation
at the Galactic center a few megayears ago \citep{krabbe95}, and that
perhaps Sgr~A* was more actively accreting mass then than it is observed
to be today.  Furthermore, there is evidence that the stellar initial
mass function near a SMBH is top heavy \citep{nayakshin05}, implying
that the Galactic center is a viable location for finding highly massive
stars.  Photometric variability is one way of probing the mixed Galactic
center stellar population on a large scale, as both evolved low mass and
high mass stars are prone to photometric variations.

There have been several photometric variability studies of the Galactic
center, some of which we summarize in Table~\ref{tbl:surveys}. While
some of these studies have covered a much longer baseline than ours does
\citep[e.g.,][]{ott99,rafelski07}, with more than 130 epochs in each
band, our light curves have a much higher sampling.  These data also
represent the first major multi-wavelength photometric variability study
of the Galactic center since \citet{blum96a}.

We present here our catalog of 110 variable stars within $\sim 2$~pc of
Sgr~A*, including 14 periodic sources.  We outline our observations in
\S~\ref{sec:obs}.  In \S~\ref{sec:analysis} we discuss the reduction of
data with image subtraction.  We discuss our variable selection in
\S~\ref{sec:selection}, and we present the catalog of variables and
their light curves along with a discussion of some of the more
well-known and interesting objects in \S~\ref{sec:catalog}.  We adopt a
Galactocentric distance of 7.6~kpc \citep{eisenhauer05}.

\section{Observations}\label{sec:obs}
The observations from which IRS~16SW was studied by \citet{depoy04} and
\citet{peeples07a} were of a field of view of $112 \times 112$
arcseconds (4.12~pc projected) approximately centered on the Galactic
center; we describe here the stellar photometric variability in these
data.  The observations of the Galactic center were made from 2000--2002
at the Cerro-Tololo Inter-American Observatory (CTIO)/Yale 1-m telescope
using the facility optical/infrared imager (ANDICAM; see
\citealt{depoy03} for details).  ANDICAM has a pixel scale of
$0.22''$~pix$^{-1}$ on a $1024\times 1024$ array.  Both $H$- and
$K$-band (centered at 1.6\um\ and 2.2\um\ respectively) images were
taken in the 2001 and 2002 observing seasons; $H$-band data were also
obtained in 2000.  The observing campaign consists of every usable night
from UTC 2000 August 13 through UTC 2000 October 14, UTC 2001 May 20
through UTC 2001 November 3, and UTC 2002 June 9 through UTC 2002
September 25.  Each night, a set of seven slightly offset images were
obtained and then combined and trimmed (decreasing the field of view) to
form a final nightly image.  The $H$-band images consist of 30~s
exposures, and it took about four minutes to obtain the group of seven
images; the $K$-band images consist of 10~s exposures and took about two
minutes to obtain.

After image quality cuts were made, there are a total of 133 $H$-band
images spanning 774 nights and 134 $K$-band images spanning 495 nights.
The full-width half-maximum (FWHM) ranges from $0\farcs 93$ to $1\farcs
93$; in general, the $H$-band images are of somewhat higher quality
(typical FWHM $\sim 1\farcs 3$) than the $K$-band (with typical FWHM
$\sim 1\farcs 45$).  The final $H$- and $K$-band images are also offset
relative to one another by $8.8''$; most of the stars detected in only
one of the two bands are due to this offset.

\section{Data Analysis}\label{sec:analysis}
\begin{figure}
\plotone{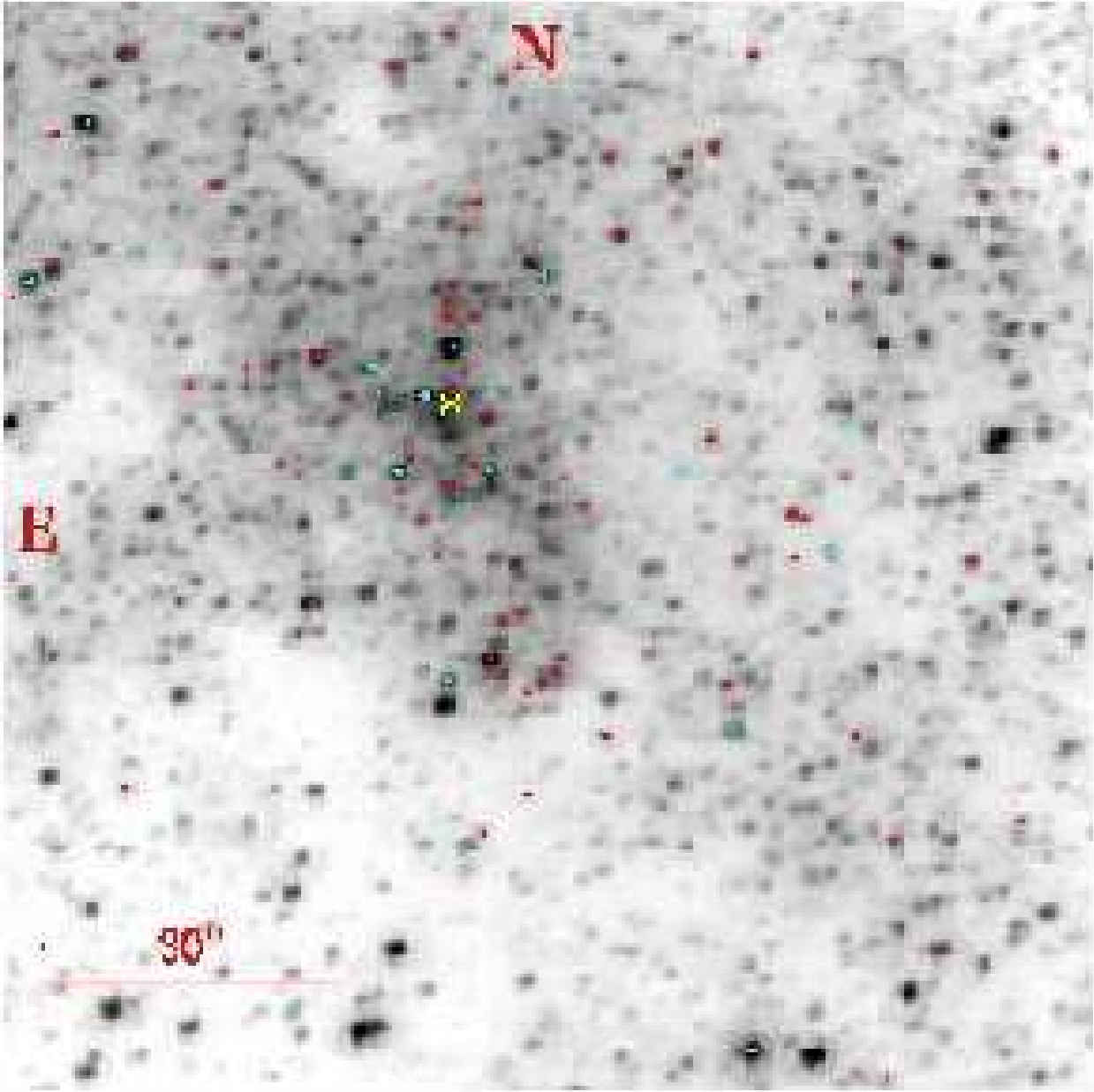} 
\caption{\label{fig:himage}The reference $H$-band image.  The red and
  cyan points mark variable sources listed in Tables~\ref{tbl:vary}
  \& \ref{tbl:periodic}; the cyan points mark stars discussed in
  \S~\ref{sec:catalog}.  The yellow X marks the location of Sgr~A*.
  North is up and East is to the left.}
\end{figure}

Because the field is crowded, we reduced the data with the ISIS image
subtraction package \citep{alard98,alard00} following the procedures
outlined by \citet{hartman04}.  ISIS measures the change in flux for an
object relative to a reference image, after correcting for the relative
change in the point spread function (PSF).  Our reference images (and
thus ``fiducial'' magnitudes) for both bands are based on images from
the 2001 observing season.  The reference $H$-band image is presented in
Figure~\ref{fig:himage}.

We identified the stars running DAOPhot/ALLSTAR progam
\citep{stetson87,stetson92} on the astrometric reference image for each
band.  The final list consists of 1665 sources in $H$-band, with $H =
9.1$ to $17.0$, and 1785 sources in $K$-band, with $K = 7.7$ to
$15.1$. We based the astrometric and photometric calibration on the
2MASS catalog \citep{strutskie06}.  For the astrometry, we obtained an
astrometric solution by matching the 2MASS $K$-band catalog to our
$K$-band data.  We similarly solved for the coordinate transformation
between the $H$- and $K$-band data sets, which enabled us to find the
coordinates of sources detected only in $H$ band.  The astrometric
solution is good to $\pm 0.24''$ in RA and $\pm 0.21''$ in dec (about
$\pm 1$~pixel in both coordinates).  We calibrated our absolute
photometry using this same set of stars matched to the 2MASS catalog;
the photometric calibration error is $\pm 0.03$~mag.

\begin{figure}
\plottwo{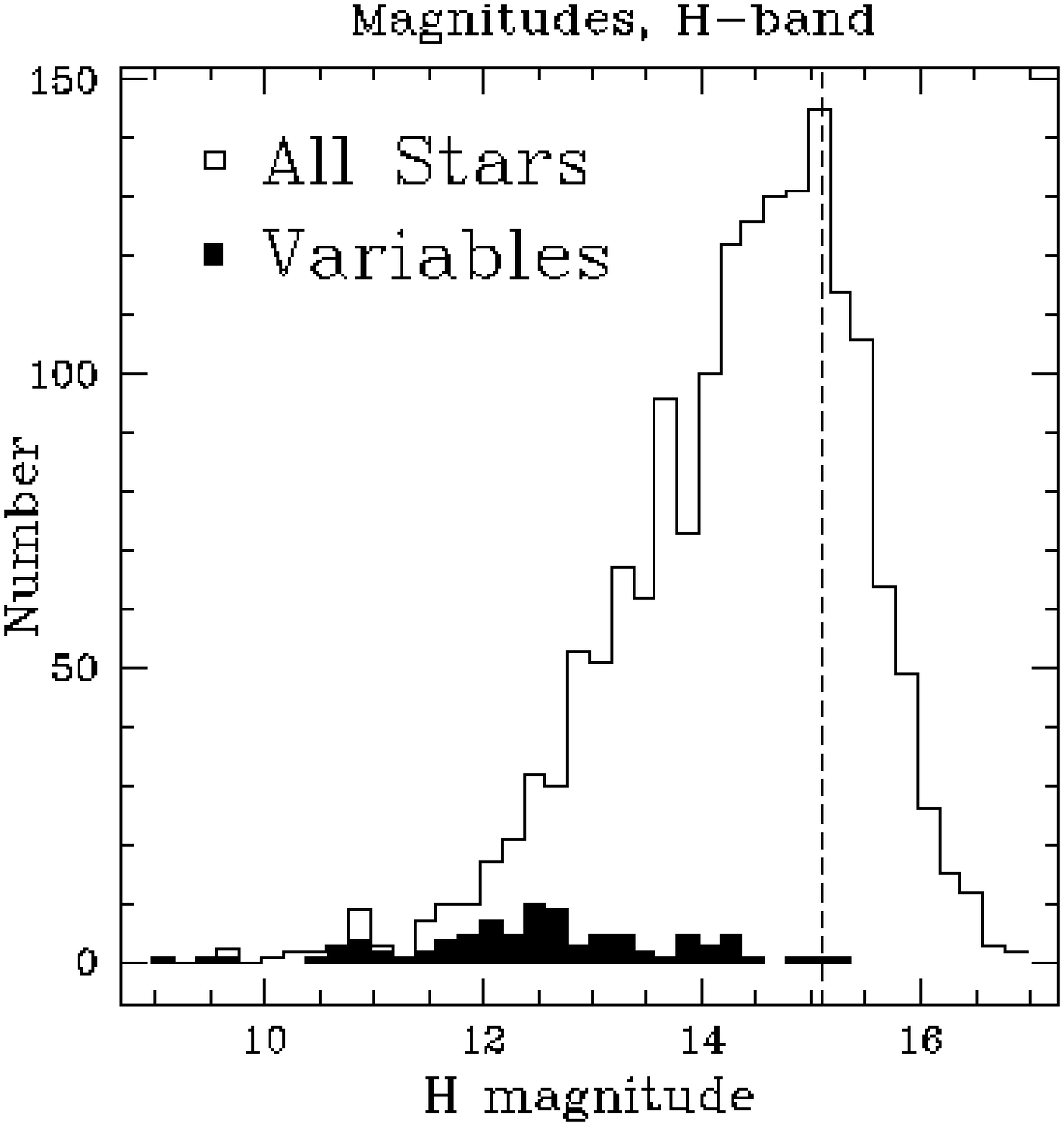}{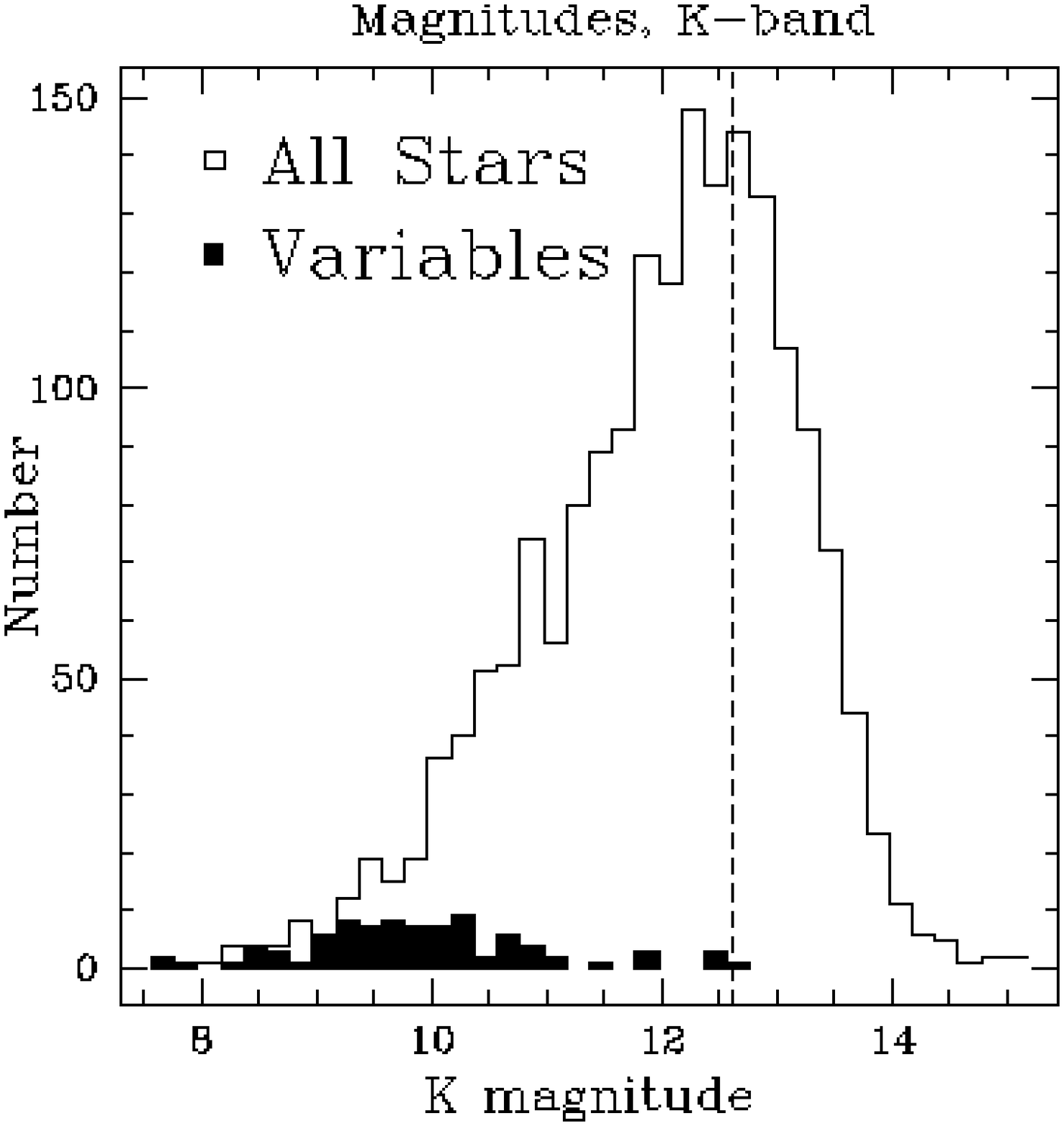}
\caption{Distribution of magnitudes in $H$ band (left) and $K$ band
  (right) for all detected sources.  The vertical dashed lines are at
  the limiting magnitude for each band (see \S~\ref{sec:analysis}).  The
  shaded histograms are the distribution of all variable sources; the
  variable sources fainter than the limiting magnitude are those sources
  which, while detected in both bands, are brighter than the limiting
  magnitude in one band, but below the limiting magnitude in the other
  band.\label{fig:mags}}
\end{figure}

\begin{figure}
\plotone{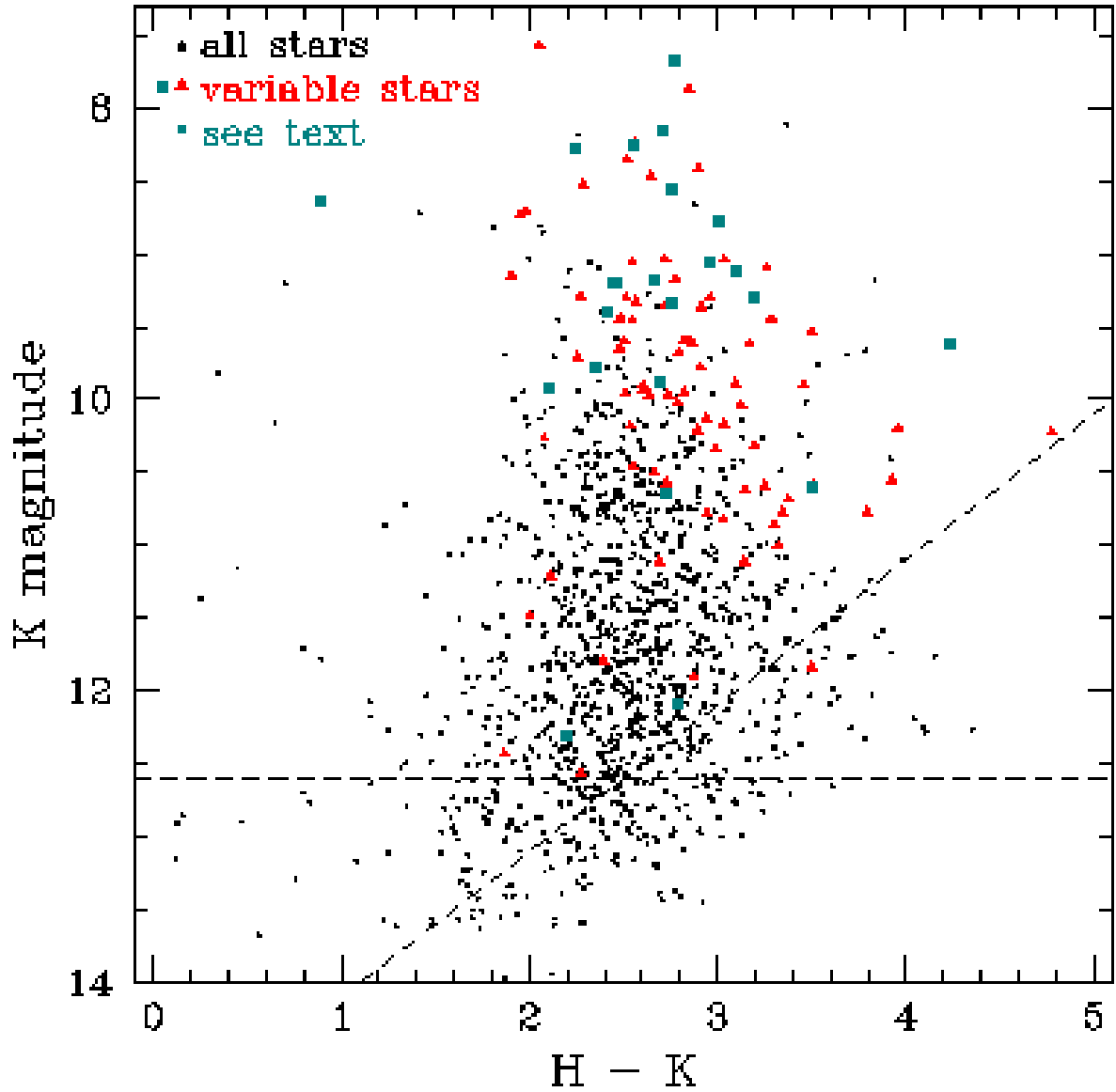}
\caption{Color-magnitude diagram.  The red triangles and the cyan
  squares are variable sources; the cyan squares are the variable
  sources discussed in \S~\ref{sec:catalog}. Most of the spread in $H-K$
  is due to differential reddening; the dashed lines correspond to the
  magnitude limits discussed in \S~\ref{sec:analysis}. \label{fig:cmd}}
\end{figure}

We define the limiting magnitude for each band to be where the magnitude
distribution peaks, as denoted by the vertical dashed lines in
Figure~\ref{fig:mags}; this is approximately the magnitude at which the
sky level starts to dominate the signal.  The limiting $H$-band
magnitude is 15.1~mag, and the limiting $K$-band magnitude is 12.6~mag;
because these magnitude determinations are based on composite images (as
opposed to individual images), the actual limiting magnitudes for
individual observations of variable sources are significantly lower.
The color-magnitude diagram is presented in Figure~\ref{fig:cmd}.

\section{Variable Selection}\label{sec:selection}
\begin{figure}
\plottwo{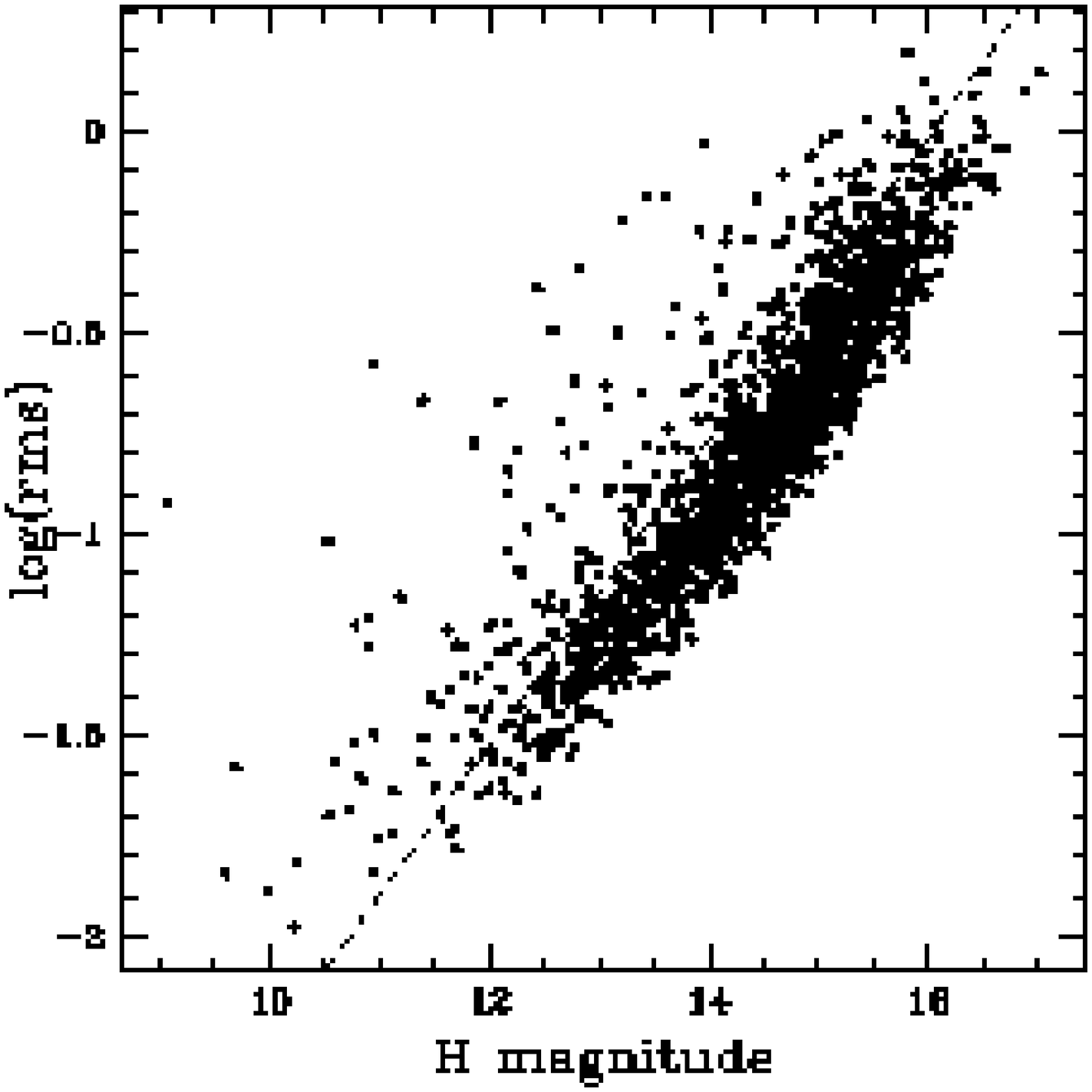}{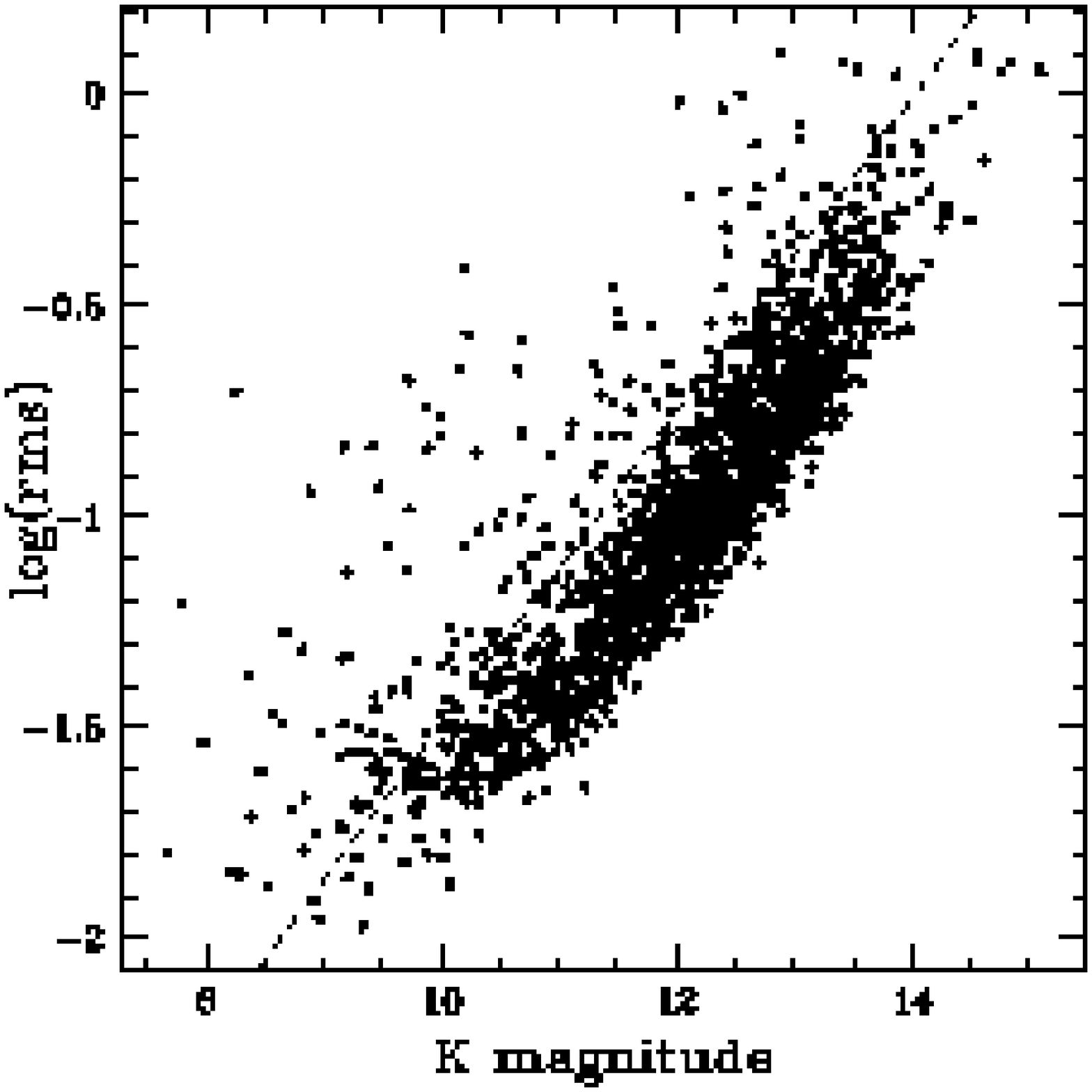}
\caption{\label{fig:rms} Plots of $\log\, \rm{rms}$ v.\ magnitude for $H$
(left) and $K$ (right).  Stars lying above the line were flagged as
``potential variables.''}
\end{figure}

The easiest to identify variable stars are those which vary about a mean
magnitude more than typical stars of the same average mangitude.  We
therefore flagged potential variables independently in both $H$- and
$K$-bands according to a root-mean-squared (rms) cut, as shown in
Figure~\ref{fig:rms}.  The cutoffs are given by
\begin{eqnarray}
\log(\rm{rms},\,H) & \ge & 0.377\, m_H - 6.044\quad\rm{and} \\
\log(\rm{rms},\,K) & \ge & 0.370\, m_K - 5.192. 
\end{eqnarray}
This led to $238/1665=14.3\%$ of $H$-band sources and $223/1785=12.5\%$
of $K$-band sources being flagged as potential variables.  Each
potential variable was visually inspected by two of us (M.S.P.\ and
K.Z.S.), and flagged as ``periodic'' (if an AoV test yielded a
believable period), ``long term,'' ``suspected,'' or not variable.  We
also performed two period searches (AoV and LS;
\citealt{schwarzenberg96,scargle82}) of all of the sources. These
searches revealed only two periodic sources not already flagged by the
rms cut as well as three other clearly variable, low-amplitude
non-periodic sources.  The final sample includes 93 sources with both
$H$- and $K$-band light curves, 14 of which are periodic, as well as 10
stars identified only in $H$-band and 7 stars identified only in $K$.
The distribution of magnitudes for all of the variable sources are shown
as the shaded regions of the histograms in Figure~\ref{fig:mags}.

Several of the sources flagged as variables were clearly blended, i.e.,
nearby stars show similar light-curve shapes.  As described in \S~4.6 of
\citet{hartman04}, the true variable is the one which displays the
greatest amplitude of variability in flux.  Thus, while in most cases we
have not attempted to correct for blending, we are able to identify
which sources are the true variables.

\section{Catalog of Variables}\label{sec:catalog}
\begin{figure}
\includegraphics[height=0.99\textheight]{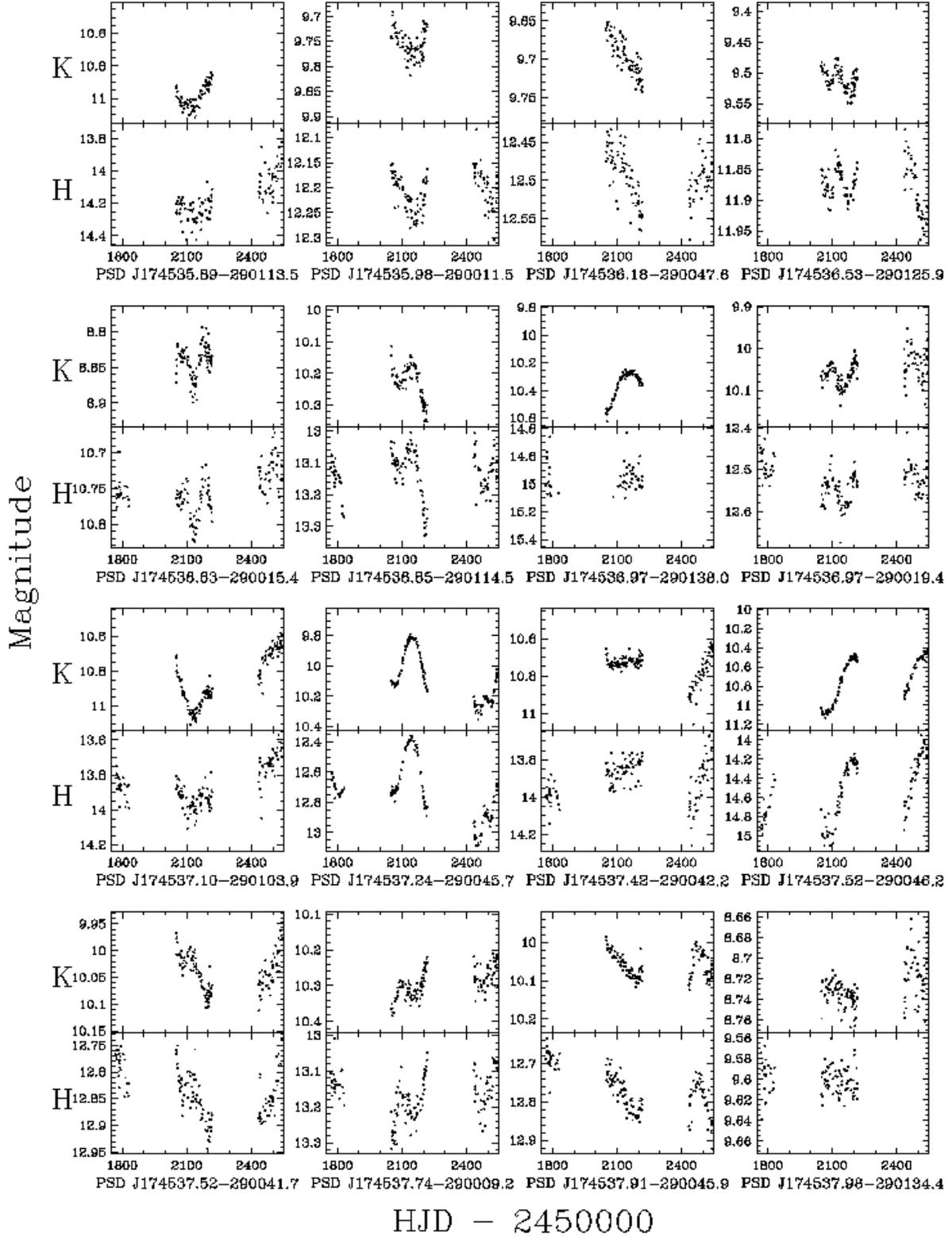}
\caption{(a)---Light curves of sources without clear period
  determinations detected in both $H$ and $K$, sorted by RA as in
  Table~\ref{tbl:vary}. For each source, the extent of the $H$- and
  $K$-band ranges plotted is the same.  Points lying below the magnitude
  limit (see \S~\ref{sec:analysis}) are not included.\label{fig:lcs}}
\end{figure}

The positions and magnitudes of our variable sources without clear
period determinations are given in Table~\ref{tbl:vary}, and the light
curves\footnote{Full light curves are available upon request from M.\
Peeples at \texttt{molly@astronomy.ohio-state.edu.}} are shown in
Figure~\ref{fig:lcs},~\ref{fig:lch}, and~\ref{fig:lck}.  The periodic
sources are summarized in Table~\ref{tbl:periodic} and
Figure~\ref{fig:periodic}.

In this section, we discuss indvidual interesting variable sources.  For
previously identified sources, we compare our data to other studies.
Table~\ref{tbl:surveys} summarizes some of the main near-infrared
variability surveys of the Galactic center to date.  At higher energies,
\citet{muno04} used the {\em Chandra X-Ray Observatory} to study X-ray
sources in the $0.5$--$8.0$~keV band in the central $9'$ of the Galaxy
from September 1999 to June 2002.  Of the more than 2000 point sources
studied, 77 sources at the Galactic center displayed long-term
variability.  On the other end of the spectrum, several masers have been
identified at the Galactic center.  Specifically, \citet{sjouwerman02}
studied 25 22~GHz H$_2$O masers within 2\dg\ and 18 43~GHz SiO masers
within $15'$ of Sgr~A*; these masers are associated with OH stellar
masers found by \citet{lindqvist92}.

A summary of the sources discussed in this section is presented in
Table~\ref{tbl:discuss}; we first discuss previously identified sources,
starting with the IRS sources (sorted numerically), followed by several
sources also found by \citeauthor{glass01}\ (sorted by their
\citealt{samus04} variable catalog name).  Finally, we discuss some of
the more interesting new variables in our catalog; these subsections are
sorted by RA, as in Tables~\ref{tbl:vary}~\&~\ref{tbl:periodic}.

\subsection{IRS~1NE}
IRS~1NE (PSD~J174540.58-290026.7) passes the rms cut described in
\S~\ref{sec:selection}, but visually does not appear to vary over the
timescales to which we are sensitive in this study.  \citet{ott99}
similarly found IRS~1NE to pass their $\chi ^2$ variability cut, but
this source did not pass their visual inspection test either.  It is
possible that IRS~1NE is in fact variable, but over shorter timescales
to which either \citeauthor{ott99}'s or our survey was sensitive, i.e.,
$\lesssim 1$~day.

\subsection{IRS~7}
IRS~7 (PSD~J174540.04-290022.7) is the brightest source in the field,
and the only one to saturate in our $K$-band images.  \citet{ott99} find
this M2 supergiant long-period variable (LPV) to have a $\Delta
K=0.26$.  We find a $\Delta H = 0.33$ (from $H = 9.0$~to~9.33).

\subsection{IRS~9}
IRS~9 (PSD~J174540.47-290034.6) is a large-amplitude variable, with
$\Delta m_H = 0.92$~mag and $\Delta m_K = 0.22$~mag (though most of the
$H$-band large amplitude variability is seen in the year 2000, when no
$K$-band data were taken).  In our data, IRS~9 is blended with two stars
detected in only $H$-band.  \citet{blum96a} found the $H-K$ color of
this M7 LPV to change by $\sim 0.2$~mag from September 1989 to July
1993. \citet{tamura96} find IRS~9 to have brightened by $\approx
0.37$~mag in $K$-band over about two years (from 1991 to 1993).
\citet{glass01} find IRS~9 (their 3-2753, also known as V4920~Sgr) to
have $\Delta m_K = 0.55$~mag over a period of 463~days; while our data
do not rule out such a period, the light curve appears more consistent
with that of a large-amplitude irregular or semi-regular variable.

\subsection{IRS~10*}
\citet{tamura96} found a variable source between IRS~10E and IRS~10W.
Dubbed\footnote{This source is referred to in the literature as
IRS~10E*, IRS~10EL, and IRS~10EE; we will use here the original name
IRS~10*.} IRS~10*, its brightness increased by at least one magnitude
between 1991 and 1992.  \citet{ott99} found similar long-period
brightness variations on the order of 1.4~mag.  Like
\citeauthor{tamura96}, we do not separately resolve IRS~10E, 10W, and
10*.  In the subtracted images from ISIS, there appears to be only one
point-source of variability in the IRS~10 region (i.e., IRS~10E and 10W
are presumably non-varying).  We associate this variability seen in the
IRS~10 complex with IRS~10*, even though DAOPhot detects two sources in
$H$-band (apparently the brighter nearby IRS~10E and 10W) and only one
source in $K$-band.  Because of this, the position we state for IRS~10*
(PSD~J174540.64-290023.6) in Table~\ref{tbl:vary} is from \citet{ott99};
the stated $K$-magnitude (10.2) is probably largely due to IRS~10E.  We
find the $K$-magnitude of IRS~10* to be relatively constant in both the
2001 and 2002 observing seasons, with the 2002 value being $\sim
0.8$~mag brighter.  Furthermore, IRS~10* was much redder in 2002 than it
was in 2001.

Both \citeauthor{tamura96}\ and \citeauthor{ott99}\ associated IRS~10*
with the OH maser OH~359.939-0.052 \citep{lindqvist92}, but
\citet{blum96a,blum96b} associated it with the \citet{lindqvist92}\ maser
OH~359.946-0.047\footnote{Though they are only two lines apart in
\citeauthor{lindqvist92}'s Table~2, OH~359.939-0.052 is roughly $35''$
further from the IRS~10 complex than OH~359.946-0.047.}.
\citet{menten97} found an SiO maser and \citet{lindqvist90} found a
H$_2$O maser coincident with OH~359.946-0.047; \citet{deguchi02} and
\citet{sjouwerman02} further found the SiO maser and the H$_2$O maser to
be variable, respectively.  \citet{muno04} additionally identify an
X-ray source, CXOGC~J174540.7-290024, within $0.8''$ of IRS~10* with an
(unobscured) X-ray luminosity $L_X \approx 1.1\times
10^{32}$~erg~s$^{-1}$.  Typically OH/IR stars are asymptotic giant branch
(AGB) stars; these pulsating stars are experiencing mass loss, which
accounts for both the NIR variability and the presence of masers.
\citet{genzel96} spectroscopically confirm that IRS~10* (their IRS~10EE)
is a late-type (i.e., consistent with being an AGB) star.

While it is possible that the X-ray flux is due to some other nearby
source (or collection of sources), under the assumption that all three
masers and the X-ray source are associated with IRS~10* (and not IRS~10E
or 10W), this might be the largest X-ray luminosity associated with an
AGB star to date \citep{karovska05}.  It has been proposed that AGB
stars should be bright X-ray sources ($L_X \sim 10^{31} -
10^{35}$~erg~s$^{-1}$) due to strong magnetic fields, but various X-ray
studies have not confirmed this prediction
\citep[e.g.,][]{soker02,kastner04a, kastner04b}; typical X-ray
luminosities for, e.g., Mira, are $\sim 10^{29}$~erg~s$^{-1}$
\citep{kastner04b}.  \citeauthor{soker02} suggest that AGB stars are
radiating in the X-ray as proposed, but this radiation is not observed
due to the high column density of the ejected material surrounding the
star (see also \citealt{blackman01}).  Furthermore, all AGB stars with
detected X-ray radiation are in binary systems (e.g., Mira~A \& B,
\citealt{karovska05}); if this trend is in fact a prerequisite for a
high AGB X-ray luminosity, then IRS~10* should have a companion.  Because
SiO, H$_2$O, and OH masers are believed to be due to different regions
of the circumstellar envelope (SiO masers are relatively close to the
star, H$_2$O masers are further out, and OH masers are even further away
from the star; \citealt{reid81}), the presence of all three
masers implies that IRS~10* is enshrouded by a substantial circumstellar
envelope.  Yet, as this material does not absorb the X-ray radiation, the
X-ray flux is more likely to be due to accretion onto a companion star
than the AGB star itself.

\subsection{IRS~12N}
\citet{tamura96}, \citet{blum96a}, \citet{ott99}, and \citet{glass01}
found IRS~12N (PSD~J174539.79-290035.2, spectral type M7III) to be
variable. \citet{tamura96} observed an increase of $\Delta m_K \sim
0.47$~mag over 2 years, and \citet{ott99} found similar variations.
\citet{glass01} found IRS~12N (their 3-2753) to be a periodic variable
of similar magnitude ($K = 8.74$) and amplitude ($\Delta m_K =
0.55$~mag) with a period $P = 463$~days.  We cannot deblend IRS~12N and
IRS~12C; five variable sources are detected within $\sim 1.5''$ of
IRS~12N with similar features to one another.  We find this source to
potentially be periodic, with $P=215$~or~$429$~days and a $\Delta K
\gtrsim 0.4$ and a $\Delta H \gtrsim 0.65$; no variations are
observed on shorter timescales.  These data are in agreement with the
classification of IRS~12N as an LPV \citep{blum96b}.

\subsection{IRS~14SW}
IRS~14SW (PSD~J174540.02-290037.2) is seen to have variability with
amplitude $\sim 0.1$~mag in both $H$- and $K$-bands in the 2001 and 2002
observing seasons.  Additionally, the $H$-band lightcurve is $\sim
0.1$~mag lower in the 2000 observing season than in 2001--2002.
\citet{tamura96} state that IRS~14SW is a ``probable'' variable;
\citet{ott99} find that IRS~14SW passes their $\chi ^2$ cut, but they do
not comment on its variability.  It is clear from our data that IRS~14SW
is, in fact, a variable source.

\subsection{IRS~14NE}
IRS~14NE (PSD~J174540.11-290036.4) is a known AGB star (spectral type
M7III; \citealt{blum96b}).  \citet{ott99} find IRS~14NE to have $m_K =
9.45$~mag and amplitude $\Delta K = 0.15$, though they do not flag it as
a potential variable.  This is consistent with our findings of $\Delta K
\approx 0.14$ and $\Delta H \approx 0.16$; we find its variability to
have clear structure.

\subsection{IRS~15SW}
IRS~15SW (PSD~J174539.99-290016.5) is a known Wolf-Rayet (WR) star (also
known as WR~101i; \citealt{vanderhucht01}) with spectral type WN8/WC9
\citep{najarro97, paumard06}.  We detect small-amplitude variations in
IRS~15SW, with $\Delta m_K \sim \Delta m_H \sim 0.15$~mag.

\subsection{IRS~16NE}
Like many of the stars in the IRS~16 cluster, IRS~16NE
(PSD~J174540.25-290027.2) is a potential luminous blue variable (LBV)
star \citep{clark05} with spectral type Ofpe/WN9 \citep{paumard06}.  We
observe IRS~16NE to have an amplitude of only $\sim 0.11$~mag in both
$H$ and $K$; furthermore, we do not observe the $H-K$ color to change
substantially or systematically.  On the other hand, the light curve is
potentially consistent with a period of 205~days.

\subsection{IRS~16SW}\label{sec:16sw}
IRS~16SW (PSD~J174540.12-290029.6) was originally proposed to be a
massive eclipsing binary by \citet{ott99}.  Using the same data
presented here, \citet{depoy04} proposed that IRS~16SW is instead a new
kind of massive pulsating star, but a recent re-reduction of the data by
\citet{peeples07a} revealed a sign error in the \citeauthor{depoy04}\
analysis.  \citet{martins06} presented a radial velocity curve of
IRS~16SW; while it appears to be a single-line binary, both their
analysis and that of \citeauthor{peeples07a}\ found the data to be
consistent with a contact binary of twin 50$\msun$ stars.  While
\citet{rafelski07} also support the idea that IRS~16SW is an eclipsing
binary, their light curve shows an asymmetry in the rise- and
fall-times---the rise-time appears to be $\sim 1.6$ times that of the
fall-time---which they propose is due to tidal deformations causing
asynchronous rotational and orbital periods of the two stars.  As
discussed by \citeauthor{peeples07a}, our data do not show such a strong
asymmetry; such an asymmetry is not seen by \citeauthor{martins06}\ either.

\subsection{IRS~28}
IRS~28 (PSD~J174540.83-290034.0) is another source described as a
``probable'' variable by \citet{tamura96}.  \citet{blum96b} classify
IRS~28 as an LPV, and \citet{glass01} find IRS~28 (their 3-72, also
known as V4923~Sgr) to be a periodic source with an amplitude of $\Delta
m_K = 0.4$~mag and a period of $P=195$~days.  We find this source to
{\em not} be clearly periodic (and a period of $\sim$200~days is ruled
out).  We find $\Delta m_K \sim 0.2$~mag and $\Delta m_H \approx
0.32$~mag; in the two observing seasons with both $H$- and $K$-band
data, the two lightcurves have similar structures, but the $H$-band
lightcurve clearly has a larger amplitude.

\subsection{IRS~34W}
\citet{trippe06} find IRS~34W (PSD~J174539.76-290026.4) to be an
irregular variable Ofpe/WN9 star, suggesting that it is a transitional
object between the O supergiant and LBV phases of its evolution.  We
find an $H-K$ color of 3.5~mag, which is consistent with what
\citeauthor{trippe06}\ found for a similar epoch.

\subsection{V4910~Sgr and V4911~Sgr}
\citet{glass01} found V4910~Sgr (their 3-270, our
PSD~J174537.24-290045.7) to have a $K=10.27$ and vary with amplitude
0.8~mag over a period of 601~days.  We find V4910~Sgr to have $H=12.6$
and $K=9.9$, with amplitudes $\Delta m_H = 0.73$~mag and $\Delta m_K =
0.56$~mag.  While a period of $\sim 600$~days is not ruled our by our
data, this source appears to be an irregular or semi-regular
large-amplitude variable with no convincing signs of periodicity.

Similarly, \citeauthor{glass01} found V4911~Sgr (their 3-88, our
PSD~J174538.02-2901002.6) to have a $K = 9.58$ with amplitude 0.5~mag
over a period of 528 days.  We find V4911~Sgr to have $H=12.5$ and
$K=9.3$, with amplitudes $\Delta m_H \approx 1.15$~mag and $\Delta m_K =
0.58$~mag (much of the variation seen in $H$-band is from the 2000
observing season, when no $K$-band data were taken).  While a period of
$\sim 528$~days is not ruled out by our data, this source appears to be
an irregular or semi-regular large-amplitude variable with no convincing
signs of periodicity.  \citet{deguchi02} associate V4911~Sgr with the
SiO maser SiO~359.930-0.045, though the projected seperation is $\sim
7''$.

\subsection{V4928~Sgr}
PSD~J174542.72-285957.4 is one of the brightest stars in our sample.
Also known as V4928~Sgr, this source is associated with OH, SiO, and
H$_2$O masers, OH~359.956-0.050\footnote{\citet{ott99} associate this
maser with IRS~10*, but they are seperated by $\approx 35''$.} at
$\alpha= 17^{\rm h}45^{\rm m}42.763^{\rm s}$, $\delta=
-28^{\circ}59'57.20''$ \citep{deguchi02, sjouwerman02}.  \citet{glass01}
find V4928~Sgr (their 3-5) to have an $m_K$ of 7.89~mag, with amplitude
$\Delta m_K = 0.65$~mag and period $P=607$~days.  We find amplitudes of
$\Delta m_H \approx 0.4$~mag and $\Delta m_K \approx 0.25$~mag, but it
is clear that the ``true'' amplitude is greater than this, as at the end
of the 2001 observing season the light curve shows no signs of
flattening.  Though a period of $\sim 600$~days does not appear
consistent with our data, we cannot rule out that V4928~Sgr is indeed
periodic.  The observed large-amplitude variability in the presence of
masers flags this source as an LPV candidate.

It is possible that V4928~Sgr is, in fact, the LPV IRS~24
\citep{blum96a,blum96b}.  The projected distance between IRS~24 (at $\alpha=
17^{\rm h}45^{\rm m}41.8^{\rm s}$, $\delta= -28^{\circ}59'59.51''$
J2000.0) and V4928~Sgr is $13.6''$; however, when \citet{levine95} first
discovered an H$_2$O maser at $\alpha= 17^{\rm h}45^{\rm m}42.7^{\rm
s}$, $\delta= -28^{\circ}59'57.2''$ J2000.0, they associated it with the
bright near-infrared IRS~24.

\subsection{V4930~Sgr}
V4930~Sgr (PSD~J174543.19-290013.0) is a known Mira (LPV) variable
\citep{samus04}.  We observe V4930~Sgr, which is one of the brighest
variable stars in our sample, to smoothly vary with an amplitude of
$\sim 1$~mag in $H$ and $\sim 0.7$~mag in $K$.  \citet{glass01} likewise
find V4930~Sgr (their 3-16) to have an amplitude of $\Delta m_K =
1.0$~mag over a period of 554~days.  While our $H$-band data do not rule
out a period of $\sim 550$~days, the light curve does not appear to be
clearly periodic.

\subsection{PSD~J174535.60-290035.4 and PSD~J174538.34-290036.7}
PSD~J174535.60-290035.4 and PSD~J174538.34-290036.7 are two of the most
enigmatic variables in our sample.  For simplicity, in this section
alone, we will refer to PSD~J174535.60-290035.4 as PSD~35.6-35.4 and
PSD~J174538.34-290036.7 as PSD~38.3-36.7.  Both of these variables are
clearly periodic. PSD~35.6-35.4 has a period $P=41.3 \pm 0.5$~days with
amplitude $\Delta m_K \approx 0.77$ and $\Delta m_H \approx 0.63$;
PSD~38.3-36.7 has a period $P=42.4 \pm 0.8$~days with amplitudes $\Delta
m_K \approx 0.66$ and $\Delta m_H > 0.57$~mag (PSD~38.3-36.7 is quite
faint in $H$-band).  Because these periods, magnitudes, and variability
amplitudes are so similar, we will discuss these two sources together.

One obvious intriguing possibility is that these stars are Cepheids.
Both of these sources have $\langle m_K \rangle \approx 12.3$~mag;
assuming these are Cepheids and taking $P=42$~days, we can calculate the
absolute $K$-band magnitude using the Cepheid period-luminosity relation
from \citet{benedict07}:
\begin{equation} \label{eqn:cepheidPL}
\langle M_K\rangle = -3.32[\log(P) - 1.0] - 5.71 = -7.78.
\end{equation}
If we now assume that these stars are at the Galactic center, with
$d_{\rm GC} = 7600$~pc, we can take this absolute magnitude and
calculate the $K$-band extinction,
\begin{equation}\label{eqn:ak}
A_K = \langle m_K\rangle - \langle M_K\rangle - 5\log d_{\rm GC, pc} + 5 = 5.68.
\end{equation}
This $A_K$, which corresponds to an $A_V = A_K/0.112 = 50.7$~mag
\citep{schlegel98}, is rather large, even for the Galactic center.
\citet{scoville03} give $A_V$ near PSD~38.3-36.7 to be $\sim 25$
(derived from Pa$\alpha$ to H92$\alpha$ radio recombination
line-emission ratios) to $\sim 34$ (derived from Pa$\alpha$ to 6~cm
radio continuum emission ratios); $A_V$ is unmeasured near
PSD~35.6-35.4.  Another way of expressing this discrepancy is to say
that if we use the \citeauthor{scoville03}'s average $A_K$ for the
field, $A_K = 3.48$~mag, then equation~(\ref{eqn:ak}) yields an $\langle
M_K\rangle$ that is $\sim 2$~mag fainter than is predicted by
equation~(\ref{eqn:cepheidPL}).

It is possible, of course, that the intervening dust is patchy on scales
smaller than \citet{scoville03} could resolve, and that one or both of
these stars happens to be behind a dense clump of obscuring material.
Under the assumption that these stars are Cepheids, we can constrain the
actual $A_K$ for each source from the observed $H-K$ colors; Cepheids
have an unreddened $\langle M_H-M_K \rangle \approx 0.0$~mag \citep{laney94}.
Taking $A_H/A_V = 0.176$ and $A_K/A_V = 0.112$ from \citet{schlegel98},
we find
\begin{equation}\label{eqn:ehk}
A_K = \frac{E(H-K)}{0.176 - 0.122}\times 0.112 = 1.75E(H-K).
\end{equation}
For $M_H - M_K = 0.0$, the observed $H-K$ color of 2.2 for PSD~35.6-35.4
yields $A_K = 3.85$~mag and the observed $H-K$ color of 2.6 for
PSD~38.3-36.7 yields $A_K = 4.55$~mag.  These extinction values are
substantially less than the $A_K = 5.68$~mag calculated above; if either
of these stars is in fact a Cepheid, some combination of a different
reddening law or a different period-luminosity relation than used here
would have to come into play.

If neither PSD~35.6-35.4 nor PSD~38.3-36.7 is a Cepheid, then the
possible nature of the variability seen in these stars is unclear.  The
shape of the light curve of PSD~35.6-35.4 remains reminiscent of a
pulsating variable; a steep brigtening is followed by a longer
fall-time.  The light curve of PSD~38.3-36.7, however, is much more
symmetric; it is possible that it is an eclipsing binary, perhaps
similar to those found by \citet{soszynski04} in the Large Magellenic
Cloud.  If this is the case, then like IRS~16SW (see
\S\S~\ref{sec:16sw}) it is both close to in contact (because the light
curve does not flatten out of eclipse), composed of near-equal surface
brightness stars (because the depths of the eclipses are the same), and
we are viewing it almost edge-on (because the depths of the eclipses are
close to 0.75~mag).  These conditions are suspicious given that, if
PSD~38.3-36.7 is a binary, then its period is 84.8~days. For comparison,
IRS~16SW has a period of 19.45~days and is comprised of two 50$\msun$
stars with a separation of $\sim 150\rsun\approx 0.7$~AU
\citep{martins06,peeples07a}.  Spectroscopic monitoring is the only way
to determine the nature of the variability of PSD~35.6-35.4 and to break
the mass-radius degeneracy if PSD~38.3-36.7 is an eclipsing binary.

\subsection{PSD~J174538.98-290007.7}
We observe PSD~J174538.98-290007.7 to have a period of $\approx
325$~days with an amplitude of $0.75$~mag in $H$-band and $0.52$~mag in
$K$-band.  Based on this large period and these large amplitudes of
variation, we propose that PSD~J174538.98-290007.7 is an LPV.

\subsection{PSD~J174537.98-290134.4 and PSD~J174539.31-290016.3}
We associate PSD~J174537.98-290134.4 and PSD~J174539.31-290016.3 with
CXOGC~J174537.9-290134 and CXOGC~J174539.3-290016 respectively
\citep{muno04}.  PSD~J174537.98-290134.4 has marginal variability, but
it is a bright source ($H = 9.5$ and $K = 8.6$).  \citeauthor{muno04}
detected neither short-term nor long-term variability for either X-ray
source.

\subsection{PSD~J174540.16-290055.7}
PSD~J174540.16-290055.7 is, within uncertainties, coincident with the
long-term X-ray variable CXOGC~J174540.1-290055 ($L_X \sim
10^{35}$~erg~s$^{-1}$, \citealt{muno04}).  We find this source to
potentially be periodic, with a period of $\approx 215$~days and an
amplitude of $\Delta m_H = 0.2$~mag and $\Delta m_K = 0.15$~mag.
\citeauthor{muno04}\ observed the X-ray flux, on the other hand, to vary
by a factor of five over the course of 17 days.  These data are
consistent with PSD~J174540.16-290055.7 being a bright cataclysmic
variable; the observed period of $\approx 215$~days is potentially the
orbital period of the evolved star around an accreting white dwarf.

\subsection{PSD~J174541.39-290126.6}
PSD~J174541.39-290126.6 is one of the reddest stars in our sample, with
an observed $H-K=4.3$.  In $H$-band we observe an amplitude of $\sim
1$~mag, though the more coherent $K$-band light curve only shows
variations of $\Delta m_K \approx 0.6$~mag.  Both \citet{wood98} and
\citet{glass01} find this object (their 61-7 and 3-205 respectively) to
have an amplitude of $\sim 0.8$--$0.9$~mag and a period of $\approx
515$~days, both of which are consistent with our data.  Furthermore,
\citeauthor{wood98} associate this large-amplitude variable with the
OH/IR star OH~359.932-0.059 \citep{sjouwerman97}.

\subsection{PSD~J174542.36-290011.2}
PSD~J174542.36-290011.2 is a periodic source with $P\sim 220$~days and
amplitudes $\Delta m_H=0.61$~mag and $\Delta m_K=0.45$~mag.
PSD~J174542.36-290011.2 shows regular brightness variations over a
period of $110.8$~days, but when phased at a period of $221.6$~days,
both the maxima and the minima are less extreme in the first half of the
oscillation than in the second. While minima alternating between shallow
and deep is characteristic of an RV~Tauri type star, a period of $\sim
220$~days is too long for PSD~J174542.36-290011.2 to be an RV~Tauri type
star \citep{sterken96}.

\subsection{PSD~J174543.14-290050.2}
We observe PSD~J174543.14-290050.2 to have amplitudes of $\Delta H \sim
0.9$ and $\Delta K \sim 0.6$.  While it does appear to be periodic, the
actual period is not clear; it is possible that the period varies from
cycle to cycle.  Based on these observations, we propose that
PSD~J174543.14-290050.2 is an LPV.

\section{Summary}\label{sec:conc}

We present a catalog of variable stars within $\sim 2$~pc of Sgr~A* in
the near-infrared $H$ and $K$ bands over a three-year baseline; $\approx
80$\% of these variables are previously unidentfied.  This is the first
photometric variability study of the Galactic center to use the
technique of image subtraction.

We find several new periodic sources.  Of these, PSD~J174535.60-290035.4
and PSD~J174538.34-290036.7 have the shortest periods, with $P\sim
42$~days.  We believe PSD~J174535.60-290035.4 is a pulsating variable,
though it is unlikely to be a Cepheid.  PSD~J174538.34-290036.7 appears
to be a nearly edge-on contact eclipsing binary system, in which case
the period is actually $84.8 \pm 1.6$~days and the individual components
are likely to be quite large.  Another periodic source,
PSD~J174540.16-290055.7 ($P\approx 215$~days), is coincident with the
X-ray variable CXOGC~J174540.1-290055; we suspect that this source is a
cataclysmic variable.  Among the previously identified sources in the
catalog, we associate IRS~10* with a bright X-ray point source and OH,
H$_2$O, and SiO masers, suggesting that it is an AGB star with an
accreting companion.

\acknowledgements{David Gonzalez and Juan Espinoza collected all the
imaging data used in this paper, and we are grateful for their
dedication and competence. The CTIO/Yale 1.0m telescope and ANDICAM are
operated by the SMARTS consortium.  We would like to thank Vincent Fish
for explaining masers and Marc Pinsonneault for explaining stars to us.
We also thank Joel Hartman for his useful \texttt{flux2mag} and
\texttt{vartools} programs, Grzegorz Pojmanski for his excellent
\texttt{lc} program, John Beacom, Andy Gould, and Rick Pogge for helpful
discusssions, and the anonymous referee for thoughtful comments.}

\clearpage

\addtocounter{figure}{-1}
\begin{figure}
\plotone{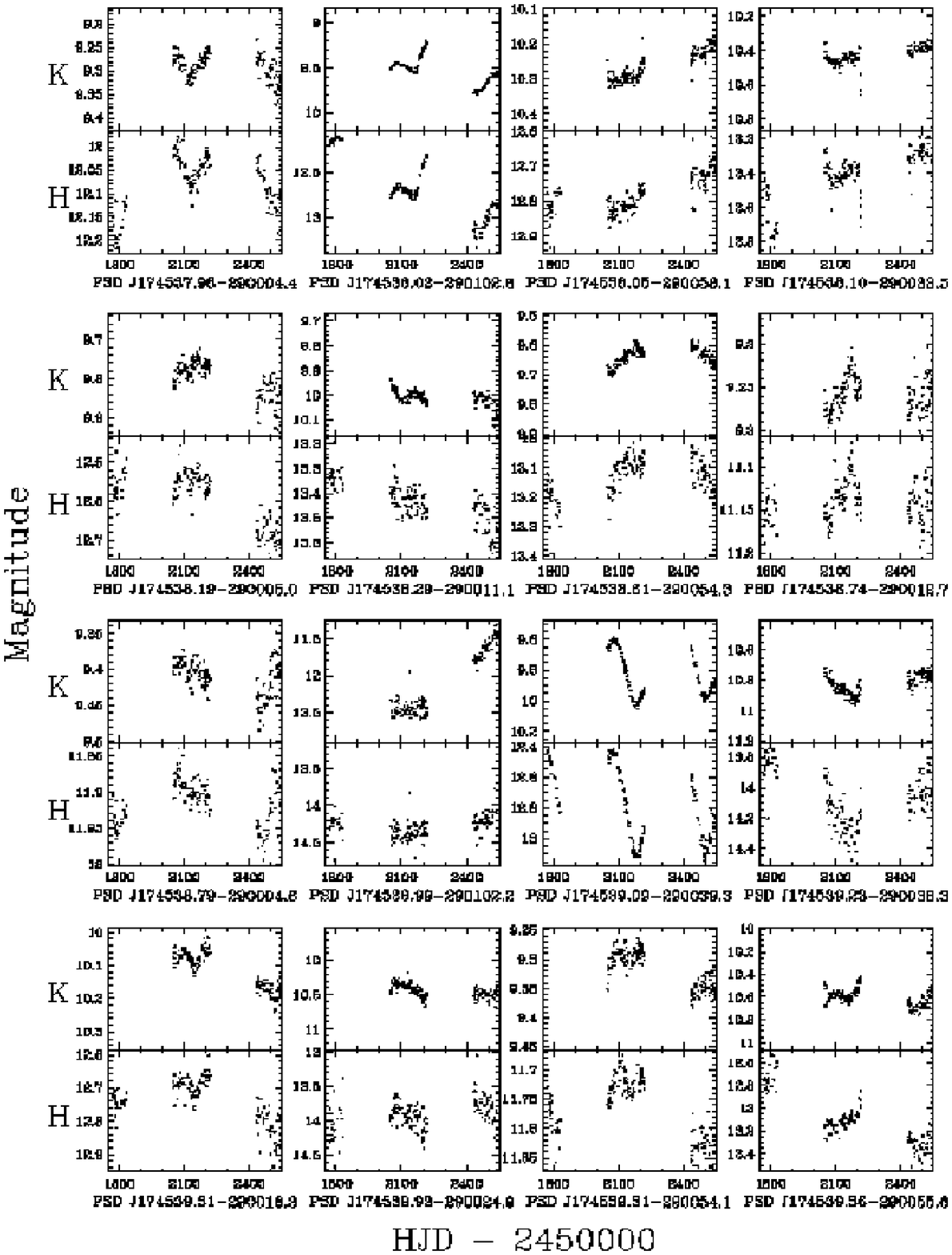}
\caption{(b)}
\end{figure}
\clearpage

\addtocounter{figure}{-1}
\begin{figure}
\plotone{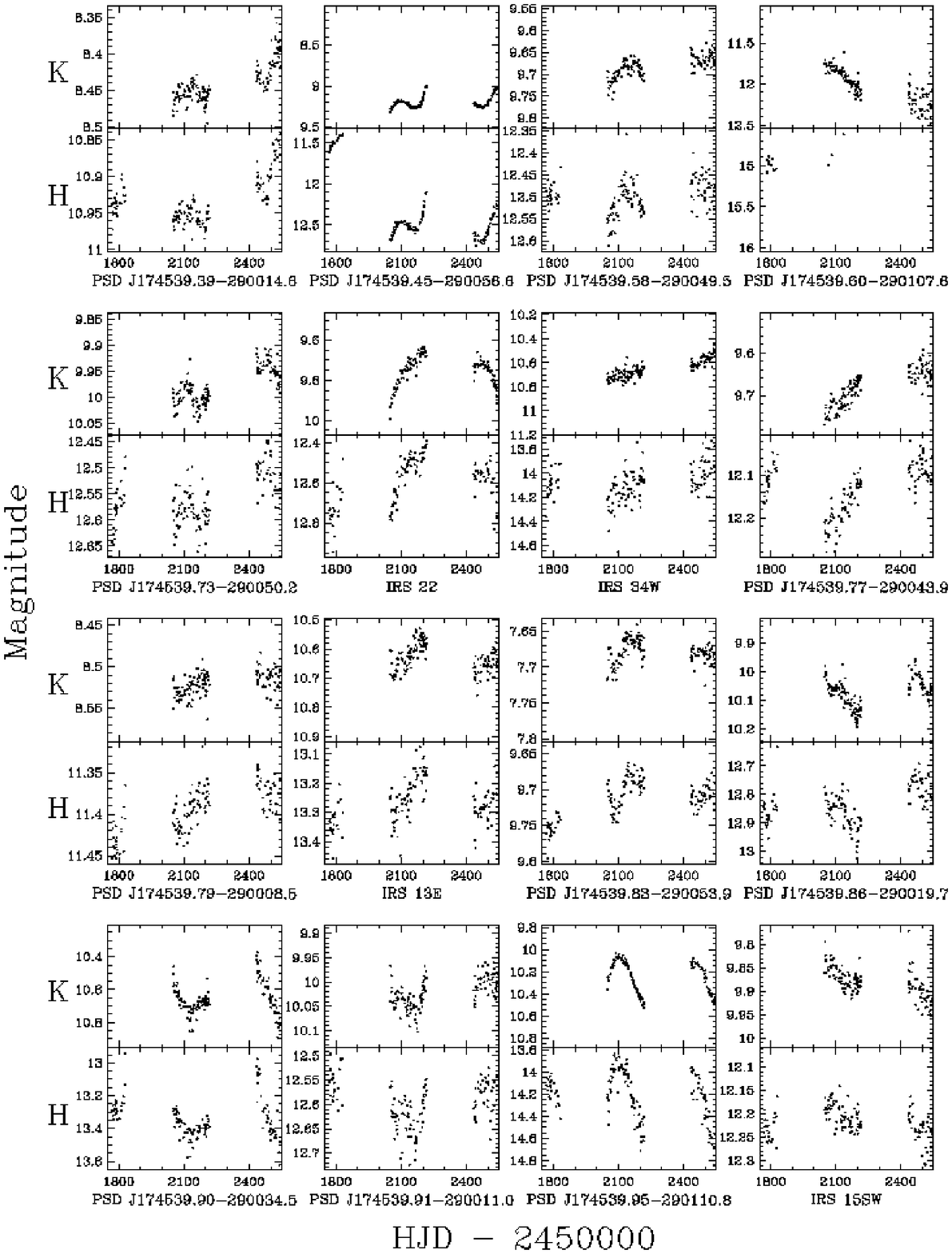}
\caption{(c)}
\end{figure}

\addtocounter{figure}{-1}
\begin{figure}
\plotone{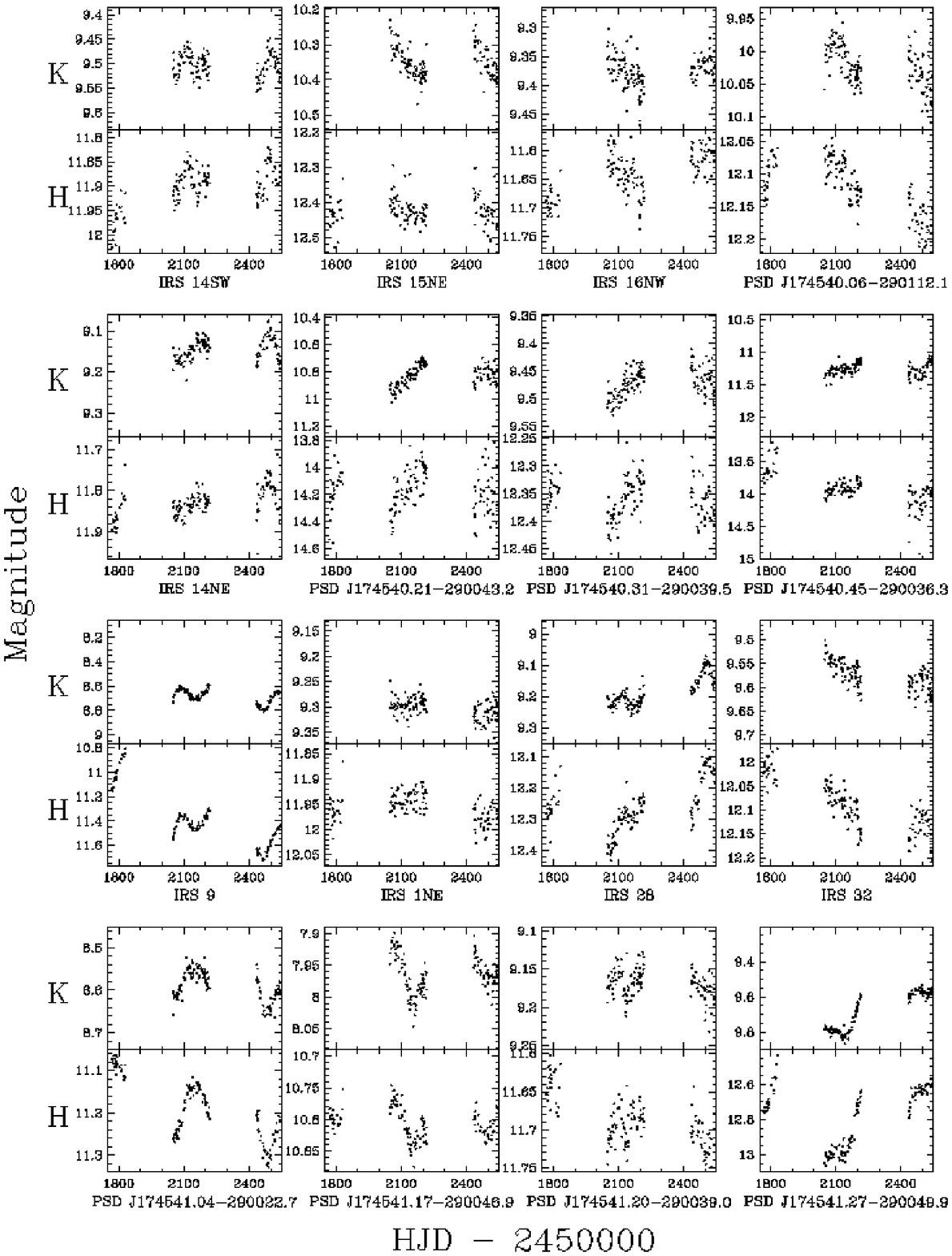}
\caption{(d)}
\end{figure}

\addtocounter{figure}{-1}
\begin{figure}
\plotone{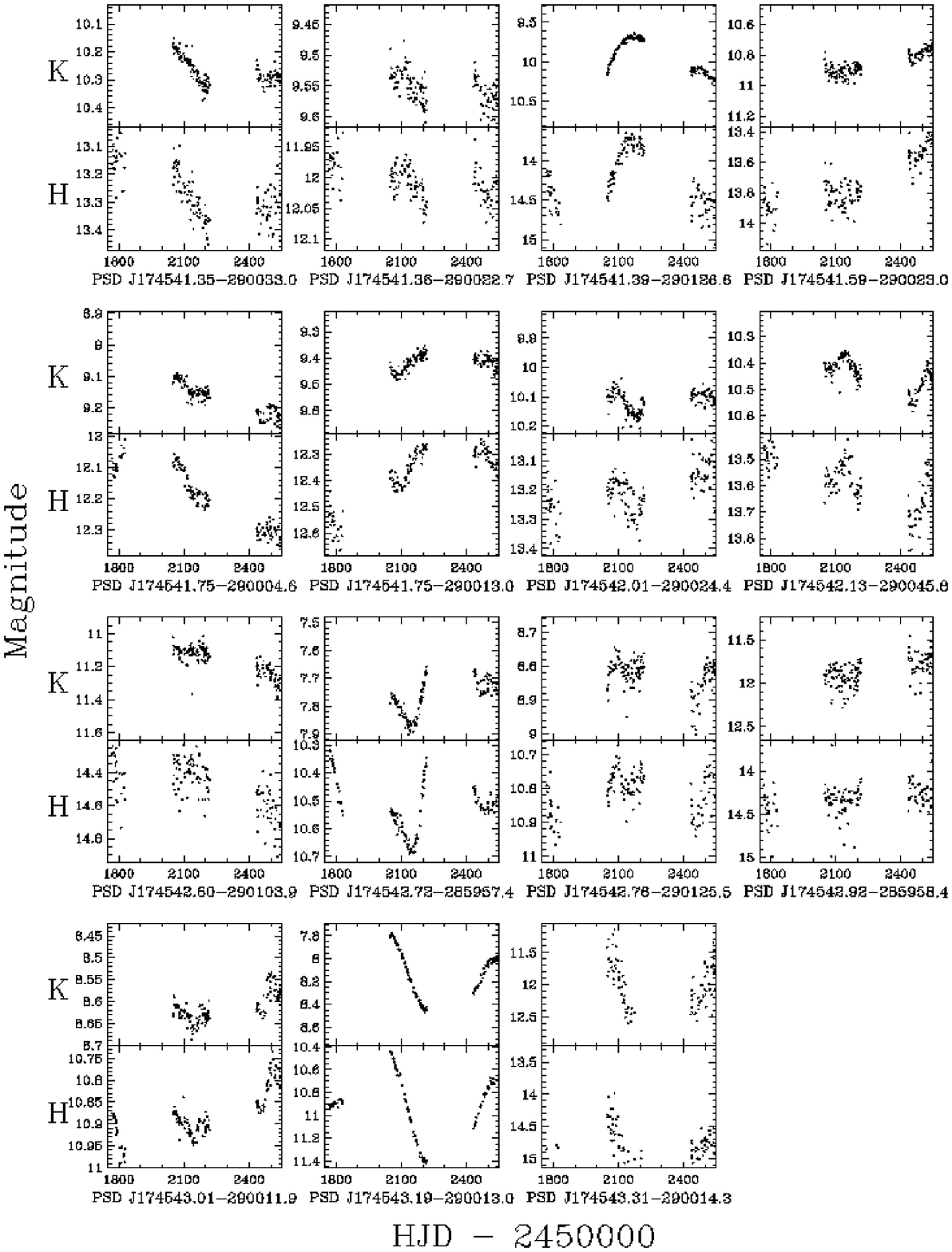}
\caption{(e)}
\end{figure}

\begin{figure}
\plotone{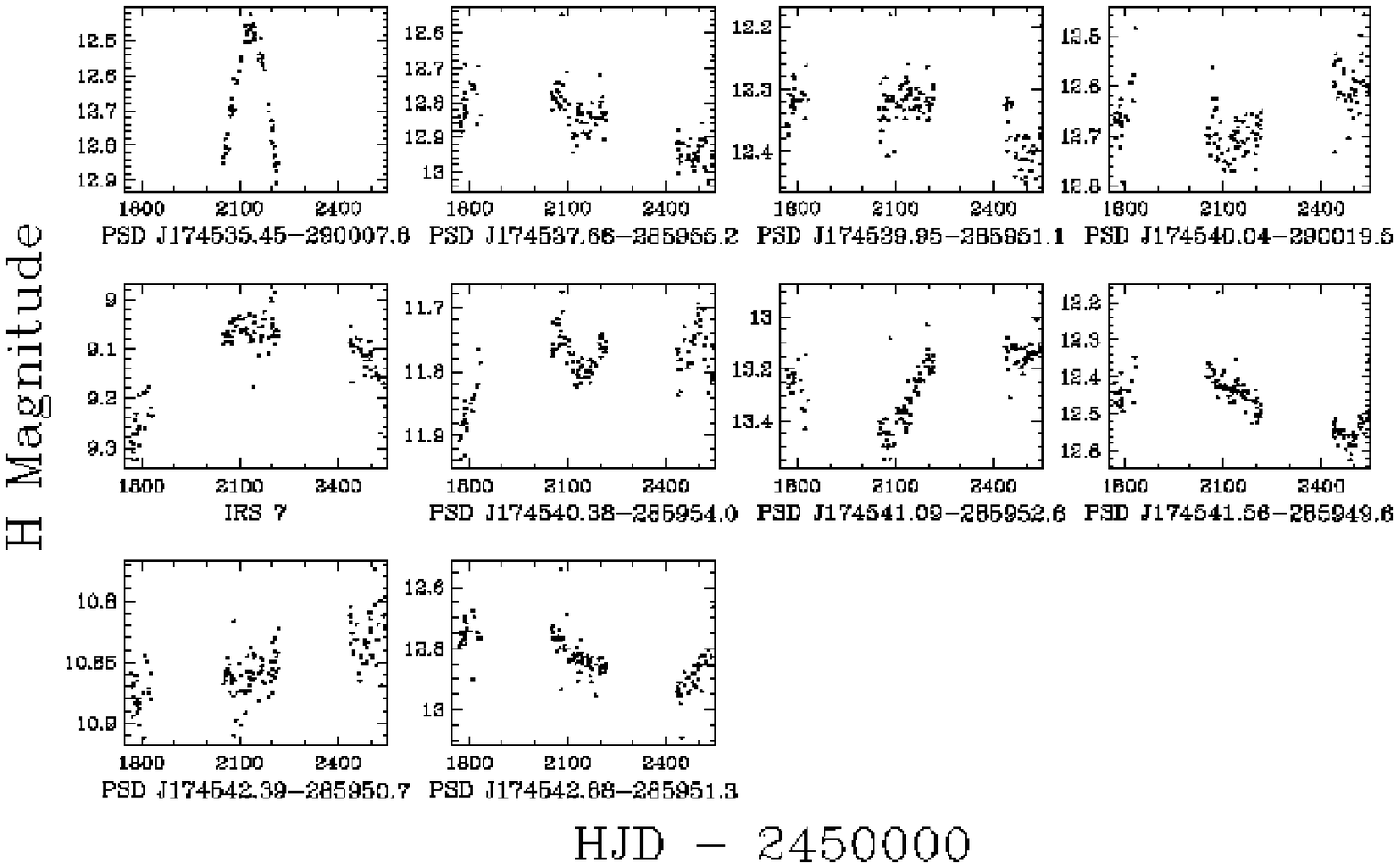}
\caption{Light curves of sources without clear period determinations
  detected only in $H$-band, sorted by RA as in
  Table~\ref{tbl:vary}. Points lying below the magnitude limit (see
  \S~\ref{sec:analysis}) are not included.
\label{fig:lch}}
\end{figure}

\begin{figure}
\plotone{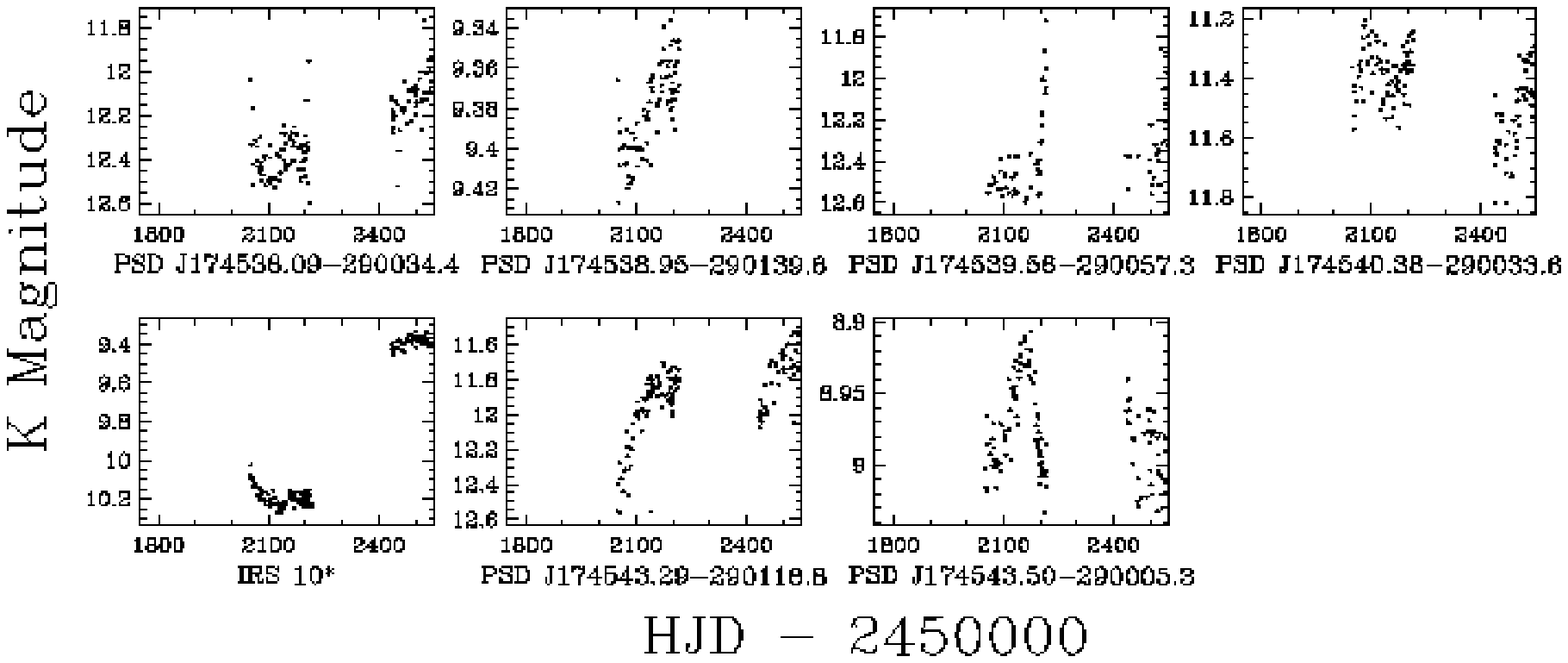}
\caption{Light curves of sources without clear period determinations
  detected only in $K$-band, sorted by RA as in
  Table~\ref{tbl:vary}. Points lying below the magnitude limit (see
  \S~\ref{sec:analysis}) are not included.
\label{fig:lck}}
\end{figure}

\begin{figure}
\plotone{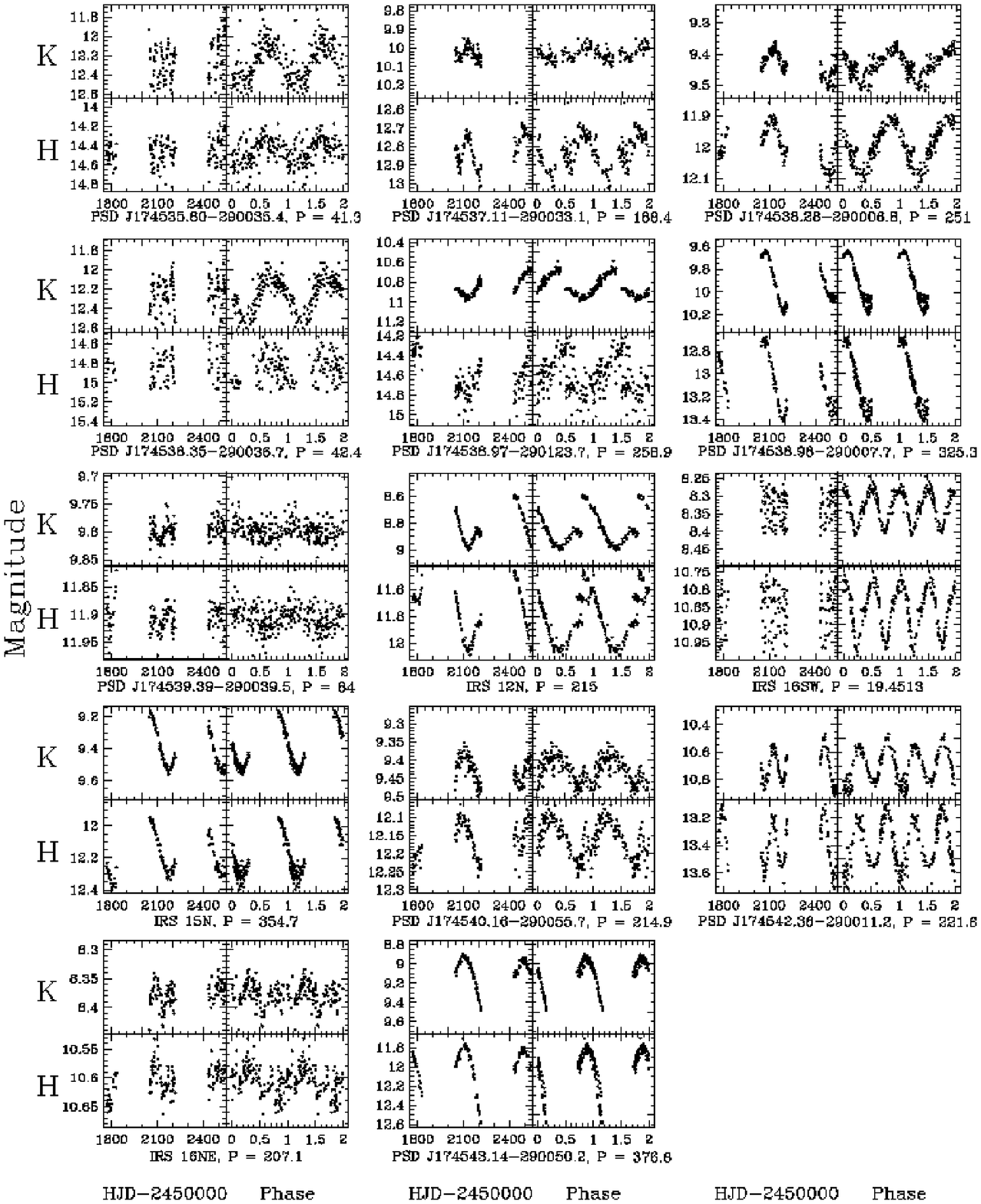}
\caption{Periodic sources.  For each source, the extent of the $H$- and
  $K$-band ranges plotted is the same; the unphased light curves are
  shown in the left panels and the phased light curves are shown in the
  right panels.  Points lying below the magnitude limit (see
  \S~\ref{sec:analysis}) are not included.  All given periods $P$ are in
  days.
  \label{fig:periodic}}
\end{figure}

\begin{deluxetable}{lrrrrl}
\tablecaption{Catalog of Variable Stars
\label{tbl:vary}  }
\tabletypesize{\footnotesize}
\tablecolumns{6}
\tablehead{
\colhead{ID} &
\colhead{$\Delta\alpha$} &
\colhead{$\Delta\delta$} &
\colhead{$H$} &
\colhead{$K$} &
\colhead{Common Names} \\
\colhead{} &
\colhead{(arcsec)} &
\colhead{(arcsec)} &
\colhead{mag} &
\colhead{mag} &
\colhead{\& Notes}
}
\startdata
PSD~J174535.45-290007.6 & $-68.9$ & $15.1 $ & 12.6 &	  &					      \\
PSD~J174535.89-290113.5 & $-62.2$ & $-50.8$ & 14.2 & 10.9 &					      \\
PSD~J174535.98-290011.5 & $-60.9$ & $11.3 $ & 12.1 & 9.7  &					      \\
PSD~J174536.18-290047.6 & $-57.9$ & $-24.9$ & 12.4 & 9.6  &					      \\
PSD~J174536.53-290125.9 & $-52.6$ & $-56.8$ & 11.9 & 9.5  &					      \\ 
PSD~J174536.63-290015.4 & $-51.2$ & $7.3  $ & 10.7 & 8.8  &					      \\
PSD~J174536.85-290114.5 & $-47.8$ & $-51.8$ & 13.1 & 10.1 &					      \\
PSD~J174536.97-290138.0 & $-46.2$ & $-75.3$ & 15.0 & 10.2 &					      \\
PSD~J174536.97-290019.4 & $-46.1$ & $3.3  $ & 12.5 & 10.0 &					      \\
PSD~J174537.10-290103.9 & $-44.1$ & $-41.2$ & 13.9 & 10.8 &					      \\
PSD~J174537.24-290045.7 & $-42.0$ & $-23.0$ & 12.6 & 9.9  & V4910~Sgr  			      \\
PSD~J174537.42-290042.2 & $-39.3$ & $-19.5$ & 13.8 & 10.6 &					      \\
PSD~J174537.52-290046.2 & $-37.9$ & $-23.4$ & 14.5 & 10.6 &					      \\
PSD~J174537.52-290041.7 & $-37.8$ & $-19.0$ & 12.8 & 10.0 &					      \\
PSD~J174537.66-285955.2 & $-35.7$ & $27.4 $ & 12.7 &	   &					      \\
PSD~J174537.74-290009.2 & $-34.4$ & $13.5 $ & 13.1 & 10.2 &					      \\
PSD~J174537.91-290045.9 & $-32.0$ & $-23.2$ & 12.7 & 10.0 &					      \\
PSD~J174537.98-290134.4 & $-30.9$ & $-71.7$ & 9.5  & 8.6  & CXOGC~J174537.9-290134		      \\
PSD~J174537.98-290004.4 & $-30.9$ & $18.3 $ & 12.0 & 9.2  &					      \\
PSD~J174538.02-290102.6 & $-30.3$ & $-39.9$ & 12.5 & 9.3  & V4911~Sgr; blended       \\
PSD~J174538.05-290058.1 & $-29.7$ & $-35.4$ & 12.7 & 10.2 &					      \\
PSD~J174538.10-290033.5 & $-29.0$ & $-10.9$ & 13.3 & 10.3 &					      \\
PSD~J174538.19-290005.0 & $-27.7$ & $17.7 $ & 12.5 & 9.7  &					      \\
PSD~J174538.29-290011.1 & $-26.2$ & $11.6 $ & 13.4 & 9.9  &					      \\
PSD~J174538.61-290054.3 & $-21.4$ & $-31.6$ & 13.0 & 9.5  &					      \\
PSD~J174538.74-290012.7 & $-19.4$ & $10.0 $ & 11.1 & 9.2  &					      \\
PSD~J174538.79-290004.6 & $-18.8$ & $18.1 $ & 11.8 & 9.3  &					      \\
PSD~J174538.95-290139.6 & $-16.3$ & $-77.0$ &      & 9.4  &					      \\
PSD~J174538.99-290102.2 & $-15.7$ & $-39.6$ & 14.3 & 12.4 &  		      \\
PSD~J174539.09-290039.3 & $-14.2$ & $-16.6$ & 12.7 & 9.8  & blended				      \\
PSD~J174539.23-290036.3 & $-12.1$ & $-13.6$ & 14.1 & 10.8 &					      \\
PSD~J174539.31-290016.3 & $-11.0$ & $6.4  $ & 11.6 & 9.2  & CXOGC~J174539.3-290016		      \\
PSD~J174539.31-290054.1 & $-11.0$ & $-31.5$ & 12.6 & 10.0 &					      \\
PSD~J174539.36-290055.6 & $-10.1$ & $-33.0$ & 13.0 & 10.5 &					      \\
PSD~J174539.39-290014.6 & $-9.7 $ & $8.0  $ & 10.9 & 8.4  & blended				      \\
PSD~J174539.45-290056.6 & $-8.8 $ & $-33.9$ & 12.4 & 9.1  & blended				      \\
PSD~J174539.56-290057.3 & $-7.1 $ & $-34.7$ &      & 12.5 &					      \\
PSD~J174539.58-290049.5 & $-6.9 $ & $-26.8$ & 12.4 & 9.6  &					      \\
PSD~J174539.60-290107.6 & $-6.6 $ & $-44.9$ & 15.3 & 11.8 & $H$ past threshold 		      \\
PSD~J174539.73-290050.2 & $-4.6 $ & $-27.5$ & 12.5 & 9.9  &					      \\
PSD~J174539.75-290055.5 & $-4.3 $ & $-32.8$ & 12.5 & 9.6  & IRS~22				      \\
PSD~J174539.76-290026.4 & $-4.1 $ & $-3.7 $ & 14.1 & 10.6 & IRS~34W			      \\
PSD~J174539.77-290043.9 & $-4.0 $ & $-21.2$ & 12.1 & 9.6  &					      \\
PSD~J174539.79-290008.5 & $-3.6 $ & $14.2 $ & 11.4 & 8.5  &					      \\
PSD~J174539.79-290029.9 & $-3.6 $ & $-7.2 $ & 13.2 & 10.6  & IRS~13E				      \\
PSD~J174539.83-290053.9 & $-3.1 $ & $-31.2$ & 9.6  & 7.6  & blended				      \\
PSD~J174539.86-290024.7 & $-2.7 $ & $-2.0 $ &      & 10.5 &				      \\
PSD~J174539.86-290019.7 & $-2.6 $ & $3.0  $ & 12.8 & 10.0 &			      \\
PSD~J174539.90-290034.5 & $-2.1 $ & $-11.8$ & 13.3 & 10.6 & blended				      \\
PSD~J174539.91-290011.0 & $-2.0 $ & $11.7 $ & 12.5 & 9.9  & CXOGC~J174539.9-290012		      \\
PSD~J174539.93-290024.9 & $-1.6 $ & $-2.3 $ & 13.9 & 10.5  &					      \\
PSD~J174539.95-285951.1 & $-1.4 $ & $31.6 $ & 12.2 &	   &					      \\
PSD~J174539.95-290110.8 & $-1.3 $ & $-48.2$ & 14.2 & 10.2 &					      \\
PSD~J174539.99-290016.5 & $-0.7 $ & $6.2  $ & 12.1 & 9.8  & IRS~15SW				      \\
PSD~J174540.02-290037.2 & $-0.3 $ & $-14.5$ & 11.8 & 9.4  & IRS~14SW				      \\
PSD~J174540.04-290018.0 & $0.0  $ & $4.7  $ & 12.3 & 10.3 & IRS~15NE				      \\
PSD~J174540.04-290022.7 & $0.0  $ & $0.0  $ & 9.0  & ---  & IRS~7, saturated in $K$		      \\
PSD~J174540.04-290027.0 & $0.0  $ & $-4.3 $ & 11.6 & 9.3  & IRS~16NW				      \\
PSD~J174540.04-290019.5 & $0.1  $ & $3.2  $ & 12.6 &	   &				      \\
PSD~J174540.06-290112.1 & $0.4  $ & $-49.4$ & 12.0 & 9.9  &			      \\
PSD~J174540.11-290036.4 & $1.0  $ & $-13.7$ & 11.8 & 9.1  & IRS~14NE				      \\
PSD~J174540.21-290043.2 & $2.6  $ & $-20.5$ & 14.1 & 10.7 &					      \\
PSD~J174540.21-290056.3 & $2.6  $ & $-33.6$ & 15.6 &	   &					      \\
PSD~J174540.31-285953.8 & $4.1  $ & $28.9 $ & 13.0 &	   &					      \\
PSD~J174540.31-290039.5 & $4.1  $ & $-16.8$ & 12.3 & 9.4  &				      \\
PSD~J174540.37-285954.0 & $5.0  $ & $28.7 $ & 11.7 &	   & blended				      \\
PSD~J174540.38-290033.6 & $5.1  $ & $-10.9$ &      & 11.4 & 			      \\
PSD~J174540.45-290036.3 & $6.2  $ & $-13.6$ & 13.8 & 11.1 & CXOGC~J174540.4-290036   \\
PSD~J174540.47-290034.6 & $6.4  $ & $-11.9$ & 11.3 & 8.6  & IRS~9, blended   \\
PSD~J174540.58-290026.7 & $8.2  $ & $-4.0 $ & 11.9 & 9.2  & IRS~1NE			      \\
PSD~J174540.64-290023.6 & $7.8  $ & $-1.4 $ &      & 10.2 & IRS~10*, blended\tablenotemark{a}        \\
PSD~J174540.83-290034.0 & $11.9 $ & $-11.4$ & 12.2 & 9.1  & IRS~28   \\
PSD~J174540.84-290027.0 & $12.0 $ & $-4.4 $ & 12.0 & 9.5  & IRS~32				      \\
PSD~J174541.04-290022.7 & $15.0 $ & $0.0  $ & 11.1 & 8.5  &				      \\
PSD~J174541.09-285952.6 & $15.8 $ & $30.1 $ & 13.2 &	   &					      \\
PSD~J174541.17-290046.9 & $17.0 $ & $-24.2$ & 10.7 & 7.9  & IRS~19, blended			      \\
PSD~J174541.20-290039.0 & $17.4 $ & $-16.3$ & 11.6 & 9.1  &			      \\
PSD~J174541.27-290049.9 & $18.5 $ & $-27.2$ & 12.8 & 9.6  &					      \\
PSD~J174541.35-290033.0 & $19.6 $ & $-10.4$ & 13.2 & 10.2 &				      \\
PSD~J174541.36-290022.7 & $19.8 $ & $-0.1 $ & 11.9 & 9.4  &					      \\
PSD~J174541.39-290126.6 & $20.3 $ & $-64.0$ & 13.9 & 9.6  &			      \\
PSD~J174541.56-285949.6 & $22.9 $ & $33.1 $ & 12.4 &	   &					      \\
PSD~J174541.59-290023.0 & $23.3 $ & $-0.4 $ & 13.7 & 10.8 &				      \\
PSD~J174541.75-290004.6 & $25.7 $ & $18.1 $ & 12.1 & 9.0  &					      \\
PSD~J174541.75-290013.0 & $25.7 $ & $9.7  $ & 12.3 & 9.3  &					      \\
PSD~J174542.01-290024.4 & $29.6 $ & $-1.8 $ & 13.2 & 10.0 &					      \\
PSD~J174542.13-290045.8 & $31.5 $ & $-23.1$ & 13.5 & 10.3 &					      \\
PSD~J174542.39-285950.7 & $35.3 $ & $32.0 $ & 10.8 &	   &					      \\
PSD~J174542.60-290103.9 & $38.5 $ & $-41.2$ & 14.3 & 11.0 &					      \\
PSD~J174542.72-285957.4 & $40.2 $ & $25.3 $ & 10.5 & 7.7  & V4928~Sgr,  blended \\
PSD~J174542.76-290125.5 & $40.9 $ & $-62.8$ & 10.7 & 8.7  &					      \\
PSD~J174542.88-285951.3 & $42.6 $ & $31.4 $ & 12.7 &	   &					      \\
PSD~J174542.92-285958.4 & $43.3 $ & $24.3 $ & 14.2 & 11.8 &					      \\
PSD~J174543.01-290011.9 & $44.6 $ & $10.7 $ & 10.8 & 8.5  &					      \\
PSD~J174543.19-290013.0 & $47.2 $ & $9.7  $ & 10.9 & 8.2  & V4930~Sgr, blended    \\
PSD~J174543.29-290118.8 & $48.8 $ & $-56.2$ &      & 11.9 &					      \\
PSD~J174543.31-290014.3 & $49.1 $ & $8.4  $ & 14.8 & 12.6 & 		     \\
PSD~J174543.50-290005.3 & $52.0 $ & $17.4 $ &      & 9.0  &					      \\
\enddata 

\tablecomments{Non-perioidic variable sources, sorted by RA.
$\Delta\alpha$ and $\Delta\delta$ are with respect to IRS~7 at
$(\alpha,\delta)=17^{\rm h}45^{\rm m}40.04^{\rm
s},\,-29^{\circ}00'22.7''$ J2000.0 \citep{blum03}.  In these
coordinates, Sgr~A* ($(\alpha,\delta)=17^{\rm h}45^{\rm m}40.04^{\rm
s},\,-29^{\circ}00'28.1''$ J2000.0) is at
$(\Delta\alpha,\Delta\delta)=(0.0'',-5.4'')$.  The magnitudes listed are
the fiducial magnitudes calculated from the ISIS reference image (see
\S~\ref{sec:analysis}); these magnitudes are {\em not} mean magnitudes.}  

\tablenotetext{a}{This position is from \citet{ott99}.}

\end{deluxetable}

\begin{deluxetable}{lrrrrrl}
\tablecaption{Catalog of Periodic Variable Stars
\label{tbl:periodic}  }
\tabletypesize{\footnotesize}
\tablecolumns{7}
\tablehead{
\colhead{ID} &
\colhead{$\Delta\alpha$} &
\colhead{$\Delta\delta$} &
\colhead{$H$} &
\colhead{$K$} &
\colhead{Period} &
\colhead{Notes} \\
\colhead{} &
\colhead{(arcsec)} &
\colhead{(arcsec)} &
\colhead{mag} &
\colhead{mag} &
\colhead{(days)} &
\colhead{} 
}
\startdata
PSD~J174535.60-290035.4 & $-66.6$ & $-12.7$ & 14.5 & 12.3 & $41.3 \pm 0.5$  & CXOGC~J174535.6-290034\\
PSD~J174537.11-290033.1 & $-44.0$ & $-10.4$ & 10.0 & 12.8 & $188.4 \pm 6.8$ &	\\
PSD~J174538.28-290006.8 & $-26.5$ & $15.9 $ & 11.9 & 9.3  & $251.0 \pm 8.4$ &		\\
PSD~J174538.35-290036.7 & $-25.5$ & $-14.0$ & 14.9 & 12.1 & $42.4 \pm 0.8$  &	 \\
PSD~J174538.97-290123.7 & $-16.1$ & $-60.9$ & 14.6 & 10.8 & $258.9 \pm 8.3$ &		 \\
PSD~J174538.98-290007.7 & $-16.0$ & $15.0 $ & 13.0 & 9.9  & $325.3 \pm 3.5$ &			 \\
PSD J174539.39-290039.5 & $-9.8 $ & $16.8 $ & 11.9 & 9.8 &  $84.0 \pm 1.3$  &   \\
PSD~J174539.79-290035.2 & $-3.8 $ & $-12.5$ & 11.8 & 8.8  & $215.0 \pm 2.4$ & IRS~12N, very blended \\
PSD~J174540.12-290029.6 & $1.2  $ & $-6.8 $ & 10.8 & 8.3  & $19.4513 \pm 0.0011$ &	   IRS~16SW  \\
PSD~J174540.13-290016.8 & $1.3  $ & $5.9  $ & 12.1 & 9.4  & $354.7 \pm 4.2$ & IRS~15N, blended \\
PSD~J174540.16-290055.7 & $1.7  $ & $-32.9$ & 12.1 & 9.3  & $214.9 \pm 2.6$ & CXOGC~J174540.1-290055   \\
PSD~J174542.36-290011.2 & $34.8 $ & $11.6 $ & 13.4 & 10.6 & $221.6 \pm 2.6$ &	   \\
PSD~J174540.25-290027.2 & $3.2  $ & $-4.6 $ & 10.5 & 8.3  & $207.1 \pm 6.1$ &	IRS~16NE	      \\
PSD~J174543.14-290050.2 & $46.5 $ & $-27.5$ & 12.0 & 9.1  & $376.8 \pm 4.1$ & blended	       \\
\enddata
\tablecomments{$\Delta\alpha$ and $\Delta\delta$ are with respect to
  IRS~7 at $(\alpha,\delta)=17^{\rm h}45^{\rm m}40.04^{\rm
  s},\,-29^{\circ}00'22.7''$ J2000.0. The magnitudes listed are the
  fiducial magnitudes calculated from the ISIS reference image (see
  \S~\ref{sec:analysis}); these magnitudes are {\em not} mean
  magnitudes. Except for IRS~16SW, the listed uncertainties on the
  periods roughly correspond to FWHM of the AoV peak (see
  \S~\ref{sec:analysis}. For IRS~16SW, the listed period and uncertainty
  are from \citet{peeples07a}; see also \S\S~\ref{sec:16sw}.}
\end{deluxetable}

\begin{deluxetable}{lllcl}
\tablecaption{Near-infrared Photometric Surveys of the Galactic Center
\label{tbl:surveys}  }
\tabletypesize{\footnotesize}
\tablecolumns{5}
\tablehead{
\colhead{Survey} &
\colhead{Field of View} &
\colhead{Time Sampling} &
\colhead{Wavelength} &
\colhead{Notes \& Variable Sources} \\
}
\startdata
\citet{tamura96} & $24'' \times 24''$ & 1991, 1992, 1993;  & $K$  & IRS~9, 10*, 12N, 14SW, 28 \\
  &                     &    3 epochs             &                   &          \\
\citet{blum96a}  & central $\sim 2'$  & May 1993--April 1995;    & $J$, $H$, $K$, $L$  & IRS 7, 9, 12N \\
  &                     &    5--8 epochs             &                   &          \\
\citet{ott99}    & $20'' \times 20''$ & August 1992--May 1998;   & $K$ & IRS 7, 9, 10*, 12N, 16SW and\\ 
  &                     &    $\sim 40$ epochs             &                   & potentially IRS 1NE, 6, 14SW         \\
\citet{glass01}  & $24' \times 24'$ & 1994--1997; $\sim 30$ epochs  & $K$  & only large-amplitude, \\ 
  &                     &                 &  & periodic variability \\
\citet{rafelski07} & $5'' \times 5''$ & 1995--2005;  50 epochs  & $K$  & IRS 16SW, 16NW, 16CC, 29N, \\ 
  &                     &                &  & and 12 S stars 
\enddata
\tablecomments{For reference, $1''$ at 7.6~kpc \citep[][Galactocentric
  distance]{eisenhauer05} corresponds to $0.0368$~pc projected; $1'$
  corresponds to 2.211~pc projected.}
\end{deluxetable}

\begin{deluxetable}{lllll}
\tablecaption{Summary of Sources Discussed in \S~\ref{sec:catalog}
\label{tbl:discuss}  }
\tabletypesize{\footnotesize}
\tablecolumns{5}
\tablehead{
\colhead{ID} &
\colhead{Figure} &
\colhead{Period} &
\colhead{Classification} &
\colhead{Alternate Name}
}
\startdata
IRS~1NE      		& \ref{fig:lcs}(d)   &  	      & 		    & PSD~J174540.58-290026.7 \\
IRS~7        		& \ref{fig:lch}      &  	      & LPV		    & PSD~J174540.04-290022.7 \\
IRS~9        		& \ref{fig:lcs}(d)   &  	      & LPV		    & PSD~J174540.47-290034.6 \\
IRS~10*      		& \ref{fig:lck}      &  	      & OH/IR, X-ray source & PSD~J174540.64-290023.6 \\
IRS~12N      		& \ref{fig:periodic} & $215.0 \pm 2.4$& LPV		    & PSD~J174539.79-290035.2 \\
IRS~14SW     		& \ref{fig:lcs}(d)   &  	      & 		    & PSD~J174540.02-290037.2 \\
IRS~14NE     		& \ref{fig:lcs}(d)   &  	      & AGB		    & PSD~J174540.11-290036.4 \\
IRS~15SW     		& \ref{fig:lcs}(c)   &  	      & WR		    & PSD~J174539.99-290016.5 \\
IRS~16NE     		& \ref{fig:periodic} & $207.1 \pm 6.1$& LBV?		    & PSD~J174540.25-290027.2 \\
IRS~16SW     		& \ref{fig:periodic} & $19.451 \pm 0.001$& binary  	    & PSD~J174540.12-290029.6 \\
IRS~28       		& \ref{fig:lcs}(d)   &  	      & 		    & PSD~J174540.83-290034.0 \\
IRS~34W      		& \ref{fig:lcs}(c)   &  	      & LBV?		    & PSD~J174539.76-290026.4 \\
V4910~Sgr    		& \ref{fig:lcs}(a)   &  	      & 		    & PSD~J174537.24-290045.7 \\
V4911~Sgr    		& \ref{fig:lcs}(b)   &  	      & SiO maser?	    & PSD~J174538.02-290102.6 \\
V4928~Sgr    		& \ref{fig:lcs}(e)   &  	      & OH/IR		    & PSD~J174542.72-285957.4 \\
V4930~Sgr    		& \ref{fig:lcs}(e)   &  	      & Mira		    & PSD~J174543.19-290013.0 \\
PSD~J174535.60-290035.4 & \ref{fig:periodic} & $41.3 \pm 0.5$ & 		    &			      \\
PSD~J174538.34-290036.7 & \ref{fig:periodic} & $42.4 \pm 0.8$ & binary??	    &			      \\
PSD~J174538.98-290007.7 & \ref{fig:periodic} & $325.3 \pm 3.5$& LPV?		    &			      \\
PSD~J174537.98-290134.4 & \ref{fig:lcs}(a)   &  	      & X-ray source	    &			      \\
PSD~J174539.31-290016.3 & \ref{fig:lcs}(b)   &  	      & X-ray source	    &			      \\
PSD~J174540.16-290055.7 & \ref{fig:periodic} & $214.9 \pm 2.6$& variable X-ray      &			      \\
PSD~J174541.39-290126.6 & \ref{fig:lcs}(e)   &  	      & OH/IR		    &			      \\
PSD~J174542.36-290011.2 & \ref{fig:periodic} & $221.6 \pm 2.6$& 		    &		      \\
PSD~J174543.14-290050.2 & \ref{fig:periodic} & $376.8 \pm 4.1$& 		    &			      \\
\enddata
\tablecomments{Sources are sorted in the order they are discussed in
  \S~\ref{sec:catalog}; all given periods are in days. }
\end{deluxetable}

\end{document}